\documentclass[10pt,journal,compsoc]{IEEEtran}

\ifCLASSOPTIONcompsoc
  \usepackage[nocompress]{cite}
\else
  \usepackage{cite}
\fi

\usepackage{graphicx}  
\DeclareGraphicsExtensions{.pdf,.png,.jpg,.jpeg}
\graphicspath{{figures/}{pictures/}{images/}{./}}

\usepackage{amsmath}
\usepackage{algorithmic}
\usepackage{array}
\usepackage{url}
\usepackage{microtype}                 % use micro-typography (slightly more compact, better to read)
\PassOptionsToPackage{warn}{textcomp}  % to address font issues with \textrightarrow
\usepackage{textcomp}                  % use better special symbols
\usepackage{mathptmx}                  % use matching math font
\usepackage{times}                     % we use Times as the main font
\usepackage{cite}                      % needed to automatically sort the references
\usepackage{tabu}                      % only used for the table example
\usepackage{booktabs}                  % only used for the table example
\usepackage[hidelinks]{hyperref}
\usepackage{comment}
\usepackage{enumitem}
\usepackage{setspace}
\newcommand{\para}[1]{\vspace{1mm}\noindent\textbf{#1}}
\usepackage{balance}

\begin{document}
\title{\LARGE PoseCoach: A Customizable Analysis and Visualization System for Video-based Running Coaching}
%
%
% author list

\author{Jingyuan Liu, Nazmus Saquib, Zhutian Chen, Rubaiat Habib Kazi, Li-Yi Wei, Hongbo Fu, and Chiew-Lan Tai% <-this % stops a space
\IEEEcompsocitemizethanks{
\IEEEcompsocthanksitem Jingyuan Liu and Chiew-Lan Tai are with Hong Kong University \protect\\ of Science and Technology. 
E-mail: jliucb@connect.ust.hk.
\IEEEcompsocthanksitem Nazmus Saquib is with Tero Labs, California, United States.
\IEEEcompsocthanksitem Zhutian Chen is with Harvard University.
\IEEEcompsocthanksitem Rubaiat Habib Kazi and Li-Yi Wei are with Adobe Research.
\IEEEcompsocthanksitem Hongbo Fu is with the City University of Hong Kong.}% <-this % stops an unwanted space
\thanks{Manuscript received xx xx, xxxx; revised xx xx, xxxx.}
}

\markboth{IEEE TRANSACTIONS ON VISUALIZATION AND COMPUTER GRAPHICS}%
{}
%{Liu \MakeLowercase{\textit{et al.}}: PoseCoach: A Customizable Analysis and Visualization System for Video-based Running Coaching}

\IEEEtitleabstractindextext{%
\begin{abstract}

Videos are an accessible form of media for analyzing sports postures and providing feedback to athletes.
{Existing sport-specific systems embed bespoke human pose attributes and {thus} can be hard to scale for new attributes, especially for users without programming experiences.}
Some systems retain scalability by directly showing the differences between two poses, {but they} might not clearly visualize the key differences that viewers would like to pursue.
Besides, video-based coaching systems often present feedback on the correctness of poses by augmenting videos with visual markers or reference poses.
However, previewing and augmenting videos limit the analysis and visualization of human poses due to the fixed viewpoints in videos, which confine the observation of captured human movements and cause ambiguity in the augmented feedback.
To address these issues, 
{we study customizable human pose data analysis and visualization in the context of running pose attributes,} such as joint angles and step distances.
Based on existing literature and a formative study, we have designed and implemented a system, \emph{PoseCoach}, to provide feedback on running poses for amateurs {{by comparing} the running poses between a novice and an expert.
\emph{PoseCoach} adopts a customizable data analysis model to allow users' controllability in defining pose attributes of their interests through our interface.
To avoid the influence of viewpoint differences and provide intuitive feedback, \emph{PoseCoach} visualizes the pose differences as part-based 3D animations on a human model to imitate the demonstration of a human coach.}
We conduct a user study to verify our design components and conduct expert interviews to evaluate the usefulness of the system.

\end{abstract}
}

\maketitle

\IEEEpeerreviewmaketitle

\IEEEraisesectionheading{\section{Introduction}\label{sec:introduction}}

\IEEEPARstart{R}{unning} is a globally popular exercise and many runners want to avoid injuries and improve their performance.
Not everyone can have access to human coaches, and thus various online materials and mobile apps have emerged to provide guidance on achieving {good} running forms.
{In sports training (including running),} 
an accessible means for novices to {adjust their postures} is to learn from pre-recorded performances of coaches or professional athletes by performing and comparing the same actions.
Despite the previous video-based systems for providing posture feedback~\cite{chen:2018:yoga,clarke:2020:ReactV}, analyzing and visualizing the differences in posture data in videos remain challenging, as discussed below.

One primary consideration in designing a video-based coaching system is the data for analysis, i.e., pose attributes to be retrieved from videos.
Parametric pose features (e.g., elbow angle) are sport-specific, such as knee bending for skiing~\cite{wang:2019:AICoach} and torso orientation for yoga~\cite{chen:2018:yoga}. Key parametric pose features to specific sports are often defined by domain experts and pre-programmed in the sport-specific systems.
Such pre-defined knowledge makes sport-specific systems hard to scale and support the analysis required by individual scenario{s}.
For example, for running, the pose attributes in question vary {with} running theories (e.g., falling forward for Pose Method~\cite{romanov:2002:PM}, stride lengths for ChiRunning~\cite{dreyer:2009:CR}).
It is thus impractical to develop a coaching system comprising all possible pose attributes to be analyzed.

\begin{figure}[!h]
 \centering
\includegraphics[width=0.72\columnwidth]{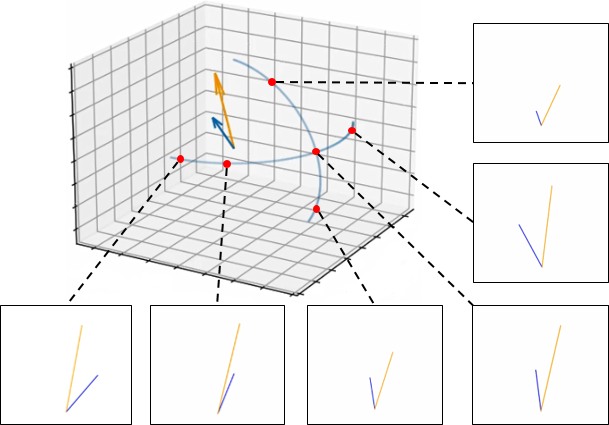}
 \caption{
 A toy example showing the ambiguity problem of 3D pose attributes.
  For a 3D angle formed by two 3D vectors (blue and orange), its appearance in 2D is largely different in the both vector lengths and the angle when observed from different viewpoints.}
 \label{fig:toyexample}
\vspace{-5mm}
\end{figure}

Besides pose attributes, another consideration is the visualization of feedback {resulting} from the analysis.
According to the taxonomy of comparison-based visualization~\cite{gleicher:2011:CIV}, existing visualizations for human pose comparison include displaying related poses in two videos side-by-side (juxtaposition)~\cite{CoachesEye:2011:CE,Hudl:2011:PAT,OnForm:2021:VAS}, overlaying one pose onto another (superposition)~\cite{clarke:2020:ReactV}, and augmenting video with visual markers (explicit encoding)~\cite{tang:2015:physio}.
A common limitation of these video-based pose comparison techniques is that the appearances of observational pose attributes, such as angles and distances, are often subject to changing viewpoints (see the toy example in \autoref{fig:toyexample}).
For sports coaching systems, such {ambiguity in the appearances of 3D attributes in videos} affects both the observation and the feedback.
Specifically, when observing the actions in videos, the 3D human pose attributes might be distorted due to perspective shortening and thus fail to reflect the actual pose attributes.
In visualization, the shapes of graphical annotation markers overlaid on videos are also subject to changing viewpoints, {failing to provide faithful feedback} to be perceived by amateur runners.
{A visualization method that accurately and intuitively shows 3D pose attributes for novices will be highly desirable for video-based coaching systems.}

% our system

\begin{figure*}[!h]
  \centering
  \vspace{-2mm}
\includegraphics[width=0.92\linewidth]{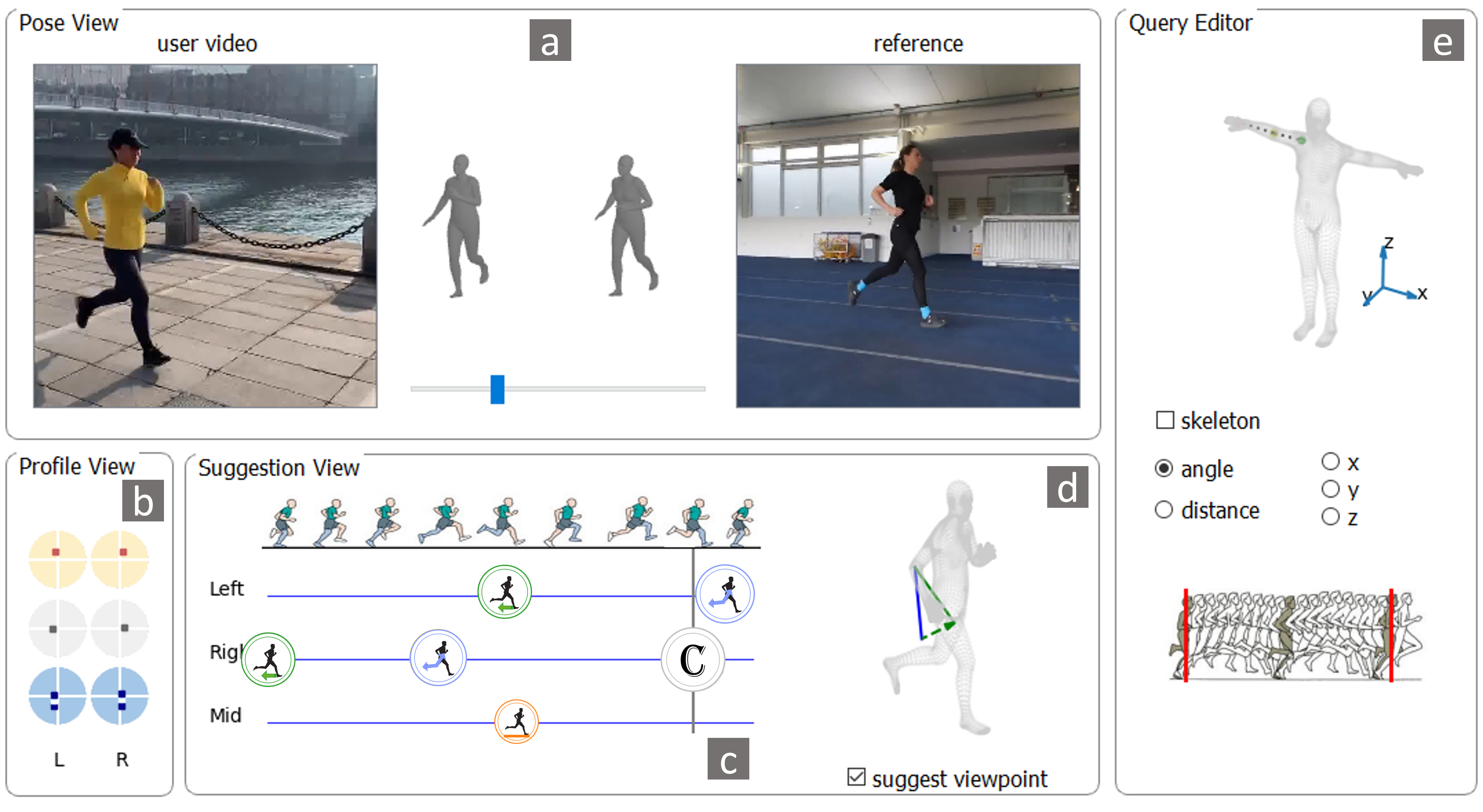}
  \vspace{-4mm}
  \caption{
  We present \emph{PoseCoach}, a novel analysis and visualization system for providing feedback on running pose for amateur runners. Our system compares two running poses (a) with respect to both common running pose attributes and specific attributes defined by users (e), and visualizes the differences with 3D animations (d).
  The profile of running poses in the {user} video (b) and differences in running poses (c) are previewed with glyphs.
  }
\label{fig:teaser}
  \label{fig:interface}
\vspace{-5mm}
\end{figure*}

In this paper, we present an interactive system for analyzing running poses from videos, aiming to address the above-mentioned issues in pose attributes and visualization.
Specifically, our goals are to both support user-defined pose attributes and provide intuitive visualizations free from the influence of the video viewpoints {so that amateur runners can easily analyze their running poses of interest (e.g., online materials they encounter)}.
To achieve these two goal{s}, we have worked closely with experts in sports science to formulate the designs of our system, \emph{PoseCoach}, based on the coaching process in practice.
As shown in \autoref{fig:teaser}, our system takes as input a {user} video containing an amateur runner's running and {a reference} video containing a professional runner's running.
{The feedback for the amateur runner is based on the comparison of the running poses in the two videos w.r.t. pose attributes, including those commonly used for evaluating running form and those customized by the {user}.}

To support customizable analysis, we design a data analysis model that allows users (advanced amateur runners or coaches) to interactively specify pose attributes to be retrieved from videos.
The data analysis model is based on a design space that unifies the parameterization of pose attributes, the representation of their differences, and users' interactions required to annotate the attributes via our interface (Figure \ref{fig:teaser}(e)).
Specifically, our data analysis model makes use of the semantic joint definition of the SMPL~\cite{Matthew:2015:SMPL} 3D human mesh model.
Users annotate and define attributes on a 3D SMPL body model in T-pose; meanwhile, the running poses in videos are also reconstructed with the SMPL model.
In this way, the annotated attributes can be retrieved from the model across multiple videos using model correspondence.
To provide intuitive feedback, we propose to visualize differences in pose attributes by animating a 3D human body model (Figure~\ref{fig:teaser}(d)), resembling the dynamic demonstration of human coaches in practice.
When previewing the 3D animation, \emph{PoseCoach} suggests viewpoints that reduce the ambiguity in visualizing 3D pose {attribute} differences caused by viewpoint.
Users can either preview with our suggested viewpoints or manually switch to other viewpoints for a better perception.

We design a user study and expert interviews to evaluate the design components and the overall effectiveness of our system.
Our main contributions can be summarized as follows:
(1) a system with a customizable data analysis model that allows interactive analysis of running poses from videos, enabling user-defined analysis instead of supporting only predefined analysis;
(2) a viewpoint-invariant visualization method for showing human pose attribute differences by 3D animation of a body model;
(3) a scheme for viewpoint suggestion that reduces ambiguity in viewpoint in previewing the pose attribute differences.

\section{Related Work}
\label{sec:prior}

\para{Pose Coaching Systems.}
Previous research work on video-based running pose analysis is limited, partly because in-the-wild running poses contain larger variations in appearance than other sports with more confined locomotion ranges, such as yoga~\cite{chen:2018:yoga} and golf~\cite{MotionPro:2018:MP}.
Running dynamics, such as ground contact time and vertical oscillation, require specific combinations of hardware to capture (e.g., \cite{Woundefinedniak:2021:MSC}).
In the following, we review posture coaching systems in general sports.

\begin{figure}[!ht]
\vspace{-1mm}
  \centering
\includegraphics[width=\linewidth]{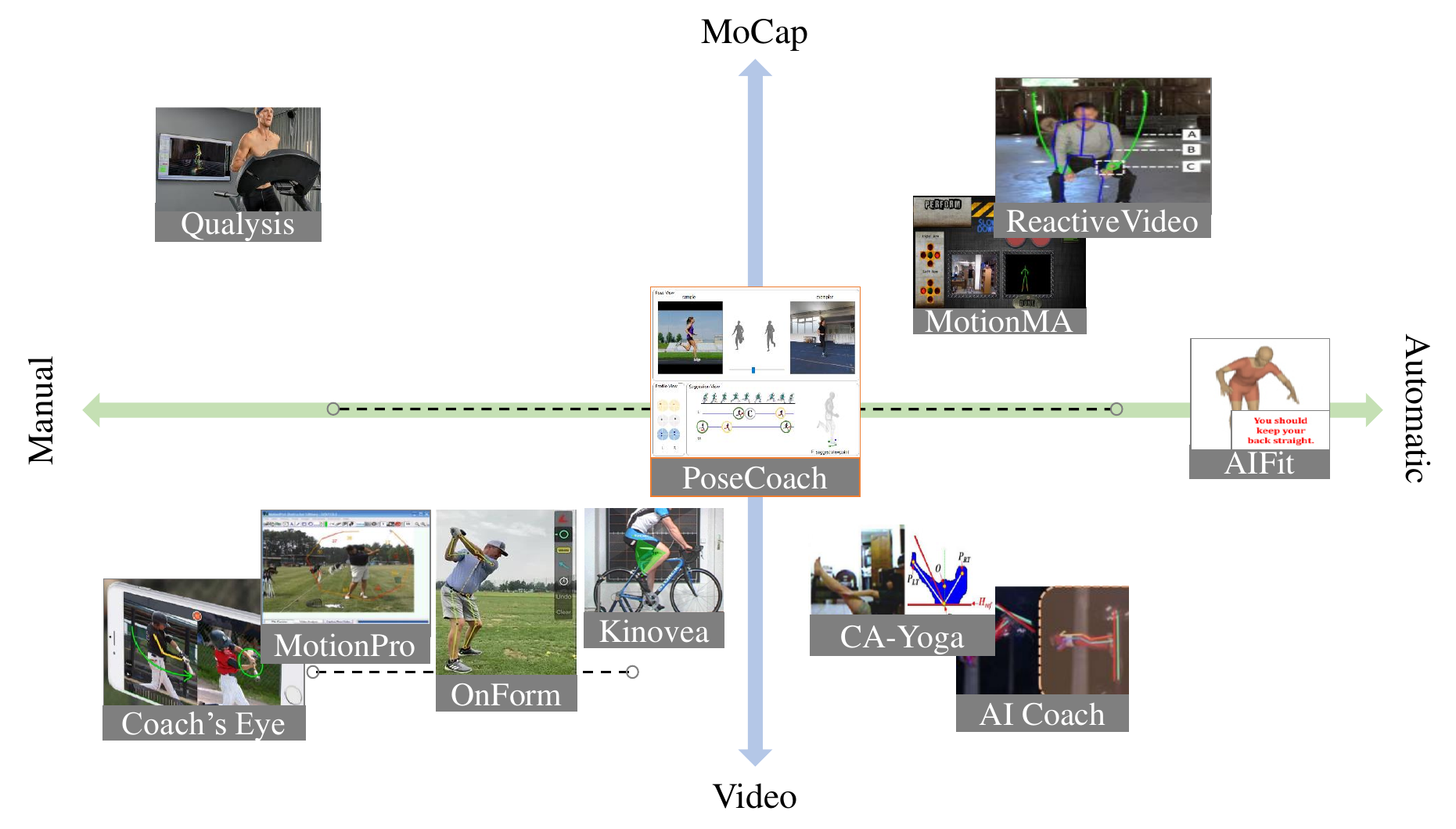}
  \vspace{-7mm}
  \caption{The spectrum of level of interference by professionals in existing coaching tools 
  (photo credits from the original sources).
  The dashed lines indicate range spans on the spectrum.
  }
\label{fig:coachtools}
\vspace{-4mm}
\end{figure}

According to how the bespoke knowledge of a specific sport is introduced into the system, existing coaching tools span the spectrum from fully-manual to fully-automatic, as illustrated in~\autoref{fig:coachtools}.
The other dimension is whether the poses are captured in 2D (videos) or in 3D (MoCap or Kinect).
The fully-manual coaching tools require human coaches to either manually annotate on video playbacks to suggest improvements~\cite{CoachesEye:2011:CE,OnForm:2021:VAS}, or analyze data of running gaits captured by MoCap~\cite{Qualysis:2015:QLS}.
MotionPro~\cite{MotionPro:2018:MP} supports manual selection of keypoints on each of the video frames such that some quantities, such as ball trajectory and 2D angles, can be obtained to facilitate analysis.
Kinovea~\cite{Kinovea:2004:AMYV} and OnForm~\cite{OnForm:2021:VAS} further simplify the manual tracking by providing basic processing of videos (e.g., automatic objects tracking and 2D human pose estimation). 

On the automatic (right) side of the spectrum, a few video-based coaching tools assess the movements based on the reconstructed 2D poses from videos using embedded rules for a specific type of sports, such as skiing (AI Coach)~\cite{wang:2019:AICoach} and yoga~\cite{chen:2018:yoga}.
Such systems would require extensive domain knowledge to design.
To avoid bespoke knowledge, some systems compute suggestions based on the comparisons between novices' actions with experts' reference actions. 
For example, MotionMA~\cite{velloso:2013:ExpNovice} and ReactiveVideo~\cite{clarke:2020:ReactV} align the experts' poses captured by Kinect onto the novices' poses in videos to visualize the difference in postures.
AIFit~\cite{fieraru:2021:AIFit} mines and highlights the most significantly different features from the comparisons of reconstructed 3D poses from videos.
Even though AIFit is fully automatic, the dominant differences might not reflect informative feedback to the sport.

{\em PoseCoach} closes the gap in both dimensions in this spectrum: 
the input is monocular videos such that it removes the constraint of indoor controlled environments, but it analyzes and visualizes in 3D to ensure spatial awareness.
It automatically performs low-level tasks but allows users the controllability to introduce high-level bespoke knowledge to the system.

\para{Sports Data Visual Analytics Systems.}
{Various visual analytics systems have been designed to analyze sports data, such as discovering winning strategies for team sports~\cite{Once:2019:OP,stein:2017:BIP,stein:2018:TSA} and racket sports~\cite{Chen:2021:ASV,wang:2022:TTVA}.
The advancement of computer vision techniques has brought about increasing opportunities for analyzing sports data in videos by tracking objects (e.g., balls) and subject{s} (e.g., players).
Video-based visual analytics systems flourish in augmenting game videos~\cite{Chen:2021:ASV,lin:2022:QOE} and instructional videos~\cite{clarke:2020:ReactV,Semeraro:2022:VIP} with visual cues to promote engagement.}
However, visual analytics systems for analyzing human poses in videos{, especially those targeting novices}, have been less explored. %focused. %, especially those target at novices.
In biomechanics stud{ies}, 3D motion analysis by reconstruction from multiview cameras has been regarded as a gold standard due to the degree of information provided~\cite{ortiz:2022:SVB}; but this often requires complex setup and expensive equipment, such as IMU sensors and force plates.
There are a few tools to facilitate {biomechanical running gait analysis in videos}, 
such as Biomechanical Toolkit~\cite{barre:2014:BT} and SkillSpector~\cite{SkillSpector:2010:SS}.
However, to reduce the ambiguity caused by viewpoint differences, the analysis {in these tools} is confined to the three anatomical planes and thus limited to indoor controlled environments.
In addition, existing video-based biomechanical analysis results are visualized with figure diagrams (e.g., angle-angle diagram and phase plane~\cite{bartlett:2014:IntroSpBio}), which {novices can hardly interpret to} obtain feedback on improvements.
To the best of our knowledge, there is no existing interactive visual analytics system for supporting in-the-wild running pose analysis, especially for novice users.

\para{User-Customizability in UI.}
Retaining user-customizability in system design has the primary advantage of promoting the user experience.
For example, several sports data visual analytics systems~\cite{lin:2022:QOE,Chen:2021:ASV} allow sports fans to select the game data of interest for visualization to facilitate a personalized game viewing experience.
{RASIPAM~\cite{wu:2022:IPM} allows experts to specify tactics for interactive explorations.}
Letting users {interact with} computational design processes~\cite{su:2018:ISN,su:2022:DIS} can also produce results tailored for users' preferences.
Besides promoting user experience, involving users in the loop has also been an effective way to retain system scalability, since new instances other than those embedded in the systems can be easily introduced without requiring end users' explicit programming.
For example, in gesture recognition, KinectScript~\cite{Nebeling:2015:KA} and Visual Gesture Builder~\cite{MSkinect:2017:VGB} allow users to interactively define gestures by recording a few repetitions.
A medical research analysis tool, DeepLabCut~\cite{mathis:2018:deeplabcut}, supports manual annotations of animals' body parts for training data-driven models to be compatible with different animal species.
Several sports coaching tools, such as MotionMA~\cite{velloso:2013:ExpNovice} and YouMove~\cite{anderson:2013:youmove}, allow users to define exemplar movements via Programming by Demonstration (PbD).
Two systems serving a similar goal to our customizable pose attributes are Kinovea~\cite{Kinovea:2004:AMYV} and RealitySketch~\cite{Suzuki:2020:ERG}, which allow users to label points or angles for tracking on top of videos.
While such keypoint definitions apply to specific videos, \emph{PoseCoach} provides a systematic set of mappings for specifying semantic human pose biomechanics that can be applied across videos.

\section{Formative Study}
\label{sec:formative}

At the beginning of this project we set out to decide the directions and the scope of a sports coaching system suitable for amateurs, which include but are not limited to runners. 
We conducted a survey on potential target users to understand their usual ways of obtaining feedback on posture correctness in practising sports (Sect.~\ref{ssec:survey}).
We also interviewed three experts on human locomotion to inform our design (Sect.~\ref{ssec:FirstInterview}).
The results of this formative study form a set of design requirements for our system (Sect.~\ref{ssec:DesignRequirements}).

\subsection{Target User Survey}
\label{ssec:survey}

\begin{figure}[!ht]
  \centering
\includegraphics[width=\linewidth]{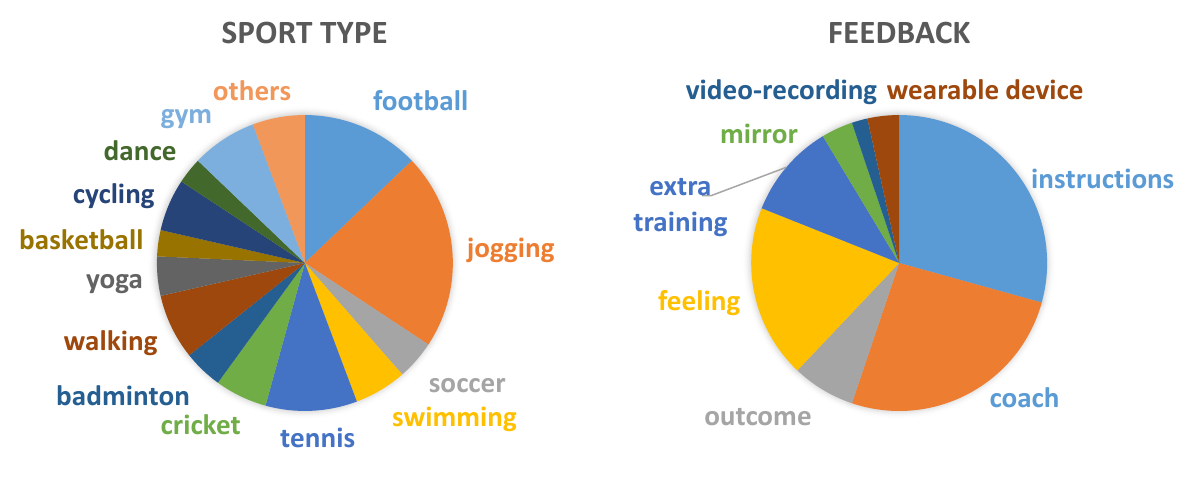}
  \vspace{-6mm}
  \caption{The visualization of survey on MTurk.
  Note that since the survey is conducted globally, we separate football and soccer according to the original responses.
  }
\label{fig:AMK}
\vspace{-3mm}
\end{figure}

To investigate the status quo and the demands of potential target users (amateur sports players) in obtaining feedback on posture during sports practice, we conducted a survey via the Amazon Mechanical Turk (MTurk).
We designed a questionnaire with three questions: (1) ``What sport(s) do you frequently practise?'' (2) ``Have you paid attention to the correctness of your body postures while practising the sport(s)?'' (3) ``If yes, please describe how you get feedback on the correctness of your postures; if not, please explain why not.''
We distributed 120 questionnaires in total, and filtered out obvious spam responses according to the quality of the short answers to question (3).
Eventually 70 effective answers were collected.
\autoref{fig:AMK} shows the summaries of responses.

Among the responses, jogging/running accounts for the most, followed by football.
Other mentioned sports include those involving posture correctness, such as yoga and swimming.
24.3\% of the subjects said they only depended on learned instructions of the actions but obtained no feedback; 21.4\% of respondents stated that they got feedback from a coach or peers.
Other main feedback includes:
5.7\% used outcome (e.g., score) as an indicator of posture correctness,
15.7\% used feeling (e.g., tense on lower back) as an indicator,
and 8.6\% adopted extra training on postures.
One respondent said he/she video-recorded the actions when practising gymnastics, and two responses explicitly said that they did not get any feedback since no one was watching.
Through this survey we learned that the public has the awareness of the importance of maintaining good postures, and there is a need for accessible posture analysis tools.
Based on the survey results, we set the focus of our system to jogging, due to its popularity and the requirement on correct postures to avoid injuries, without needing to consider ball trajectories for instrument sports or tactics for team sports.

\subsection{Expert Interviews}
\label{ssec:FirstInterview}

In order to understand the process and the key factors of human movement analysis, we conducted semi-structured interviews with three experts, two were medical doctors in Sports Medicine working in a hospital (\textbf{E1}, \textbf{E2}), and the other one (\textbf{E3}) was a researcher in sports science in a startup company studying performance analysis in sports.
During the interviews we first invited the participants to describe a representative case in which human movement analysis is involved in their daily practice. 
During the description, they were asked to identify what is the routine they analyze human movements, what are the key factors they focus on, and what is the decision process based on their observations. 
Then we raised open questions such as difficulties in human movement analysis, and the role of video-based analysis in practice.

All of the three experts mentioned that human movement analysis is based on gold standards, i.e., comparisons with the normal values in rehabilitation exercises or with top athletes' postures and performances in sports. 
Even for a full-body movement only a few key factors are concerned in evaluation (deterministic models~\cite{hay:1978:BST}).
For example, \textbf{E1} described a case of imbalance testing, where the key factors were movement accuracy and time required for completion.
\textbf{E3} emphasized the advantage of externally-focused training over internally-focused training~\cite{wulf:2010:EVI}. 
He pointed out that even though real-time feedback provides direct guidance, it would distract a subject during the action by {interfering with} the subject's intention of movements.
In addition, since a coach's attention is limited, he/she often can only focus on a specific body part during instruction, and that it would be ideal to analyze other parts during playback.
He also stated that the optimal running form for each individual {might} %may 
vary due to the differences in body configurations.
Specifically, athletes with similar heights, leg lengths, and training years are likely to have similar optimal running poses.
In practice, {professional} athletes adjust their performances ({including} but not limited to postures) by shortening the gap with a chosen exemplar, which is an elite athlete in their fields (``elite athlete template''~\cite{bartlett:2014:IntroSpBio}).
{Such a strategy also applies to amateur runners' training, where they are recommended to compare to an exemplar as a reference, and switch to a different exemplar if discomfort arises.}

Since our system is focused on running, throughout the project we closely worked with \textbf{E3} and another expert (\textbf{E4}), a third-year postgraduate student in sports science, who was involved after this formative study.
We initiated discussions with them as needed via remote chats.

\subsection{Design Requirements}
\label{ssec:DesignRequirements}

From the expert interviews on the human movement analysis, as well as the limitations of existing systems, we identify the following design requirements:

\begin{description}

\item[R1 - Show feedback as pose differences instead of correctness.]

As mentioned by \textbf{E3}, there is no absolute ``correct'' running pose.
We thus adopt the idea of the ``elite athlete template''~\cite{bartlett:2014:IntroSpBio} and design a comparison-based system.
Thus our system aims to {show users differences between} two poses and inform {them of} the gap from the current ``elite athlete template,'' instead of classifying a pose as correct or not based on predefined criteria.
Since potential reference videos exist in large volumes from the Internet, we postulate that users are responsible for finding suitable reference videos (e.g., containing a proper reference professional runner running at speed suitable to their current levels) for comparison.

\item[R2 - The comparison should be robust to variations in videos.]
Following the design requirement \textbf{R1}, the videos input by users for comparison may contain large variations in the running poses, due to background, viewpoint, subject's physical characteristics, running speed, etc.
The data analysis model should thus include pre-processing on videos to factor out {such} interference and retains only factors that {reflect the} running form, such as posture characteristics and timings.
Our system makes {few} assumptions about the qualifications of {subjects} in the reference videos; instead, it ensures {the} robustness of its comparisons in the presence of the variations in videos.

\item[R3 - The visualization should show part-based differences.]
As pointed out by \textbf{E3}, the attention of both coaches and athletes is limited, they are often advised to correct one part at a time.
Thus instead of showing all the mistakes at the same time, our system should show the differences in each body part separately.
\textbf{E3} also mentioned that for both coaches and athletes the quantitative figures do not make sense; they desire a direct corrective suggestion.
Thus instead of presenting analysis results as infographics, we need to design an intuitive way to demonstrate the differences.

\item[R4 - The system should enable user interactivity.]

To address the limitation{s} of existing systems in supporting only the analysis of predefined running pose attributes, our system should allow users to interactively define what pose attributes to be retrieved from videos.
As suggested by \textbf{E4} in a later discussion, when a coach corrects an action, he/she usually first points out the mistakes, and then shows the correct action.
Our system should also follow this routine.
Since there is no remote coach explaining the results, the visualization of comparison results should be intuitive and easily understandable by novices. 
Our system should allow users to explore the feedback to make the most sense out of it.

\end{description}

\section{System Architecture}
\label{sec:system}

In this section, we first give an overview of system components in \emph{PoseCoach}.
Then we introduce the data in Sect.~\ref{ssec:mappings} and the customizable data analysis model in Sect.~\ref{sec:method}.

\vspace{-2mm}
\subsection{System Overview}

We design our system \emph{PoseCoach} based on the aforementioned requirements.
Since we target novice users, the overall system workflow follows the ``overview first, details-on-demand'' principle~\cite{shneiderman:2003:EHI}.
Users input videos and preview suggestions through the user interface (\autoref{fig:interface}).
The input to our system contains two videos (\autoref{fig:interface}(a)): a user video to be analyzed, and a reference running video for comparison (\textbf{R1}).
Upon loading the two videos, our system automatically processes the videos to reconstruct 3D human poses, normalizes the motions (\textbf{R2}), and segments the videos into running cycles.
Our system then performs the pose analysis by aligning the running pose sequences in the user and the reference videos based on 3D pose similarity, and retrieves the pre-defined key attributes to conduct comparisons.
The suggestions for correction are generated based on the part-based differences from the comparison (\textbf{R3}), and directly reflect on a timeline tailored for running pose sequences (\autoref{fig:interface}(c)).
Those attributes that require improvement are represented with glyphs.
By clicking on each glyph on the timeline (\textbf{R4}), a detailed instruction for improving the corresponding attribute is shown as a short 3D animation of a body part on a human model in the suggestion preview window (\autoref{fig:interface}(d)).
Users can rotate the body model to navigate through viewpoints for better perception (\textbf{R4}).

\begin{figure}[!ht]
  \centering
  \vspace{-2mm}
\includegraphics[width=0.98\linewidth]{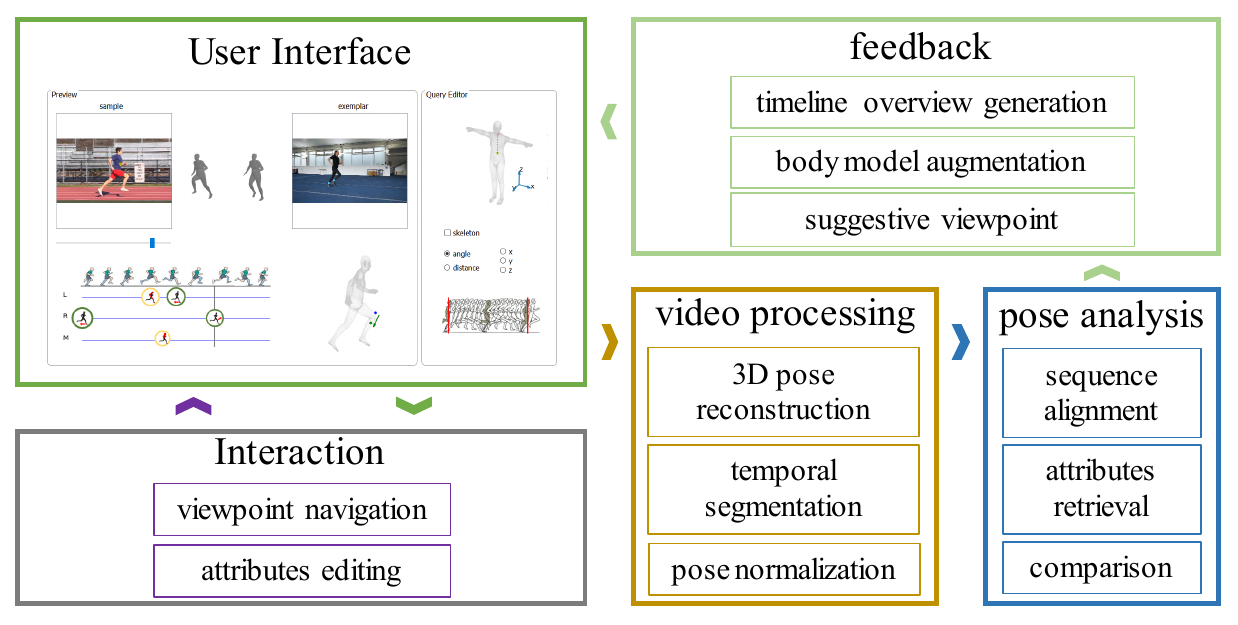}
\vspace{-3mm}
  \caption{The architecture of {\em PoseCoach}, which comprises five main modules.}
\label{fig:architecture}
\vspace{-4mm}
\end{figure}

For other pose attributes that are not embedded in our system as pre-defined attributes, the users can interactively label (\textbf{R4}) on a 3D body model via the query editor (\autoref{fig:interface}(e)).
The labeled attributes will then be retrieved and analyzed from the videos in the same way as the pre-defined attributes.
Figure~\ref{fig:architecture} shows the five modules of our system.

\subsection{{Data}}
\label{ssec:mappings}

\begin{figure*}[!ht]
  \centering
\includegraphics[width=0.98\linewidth]{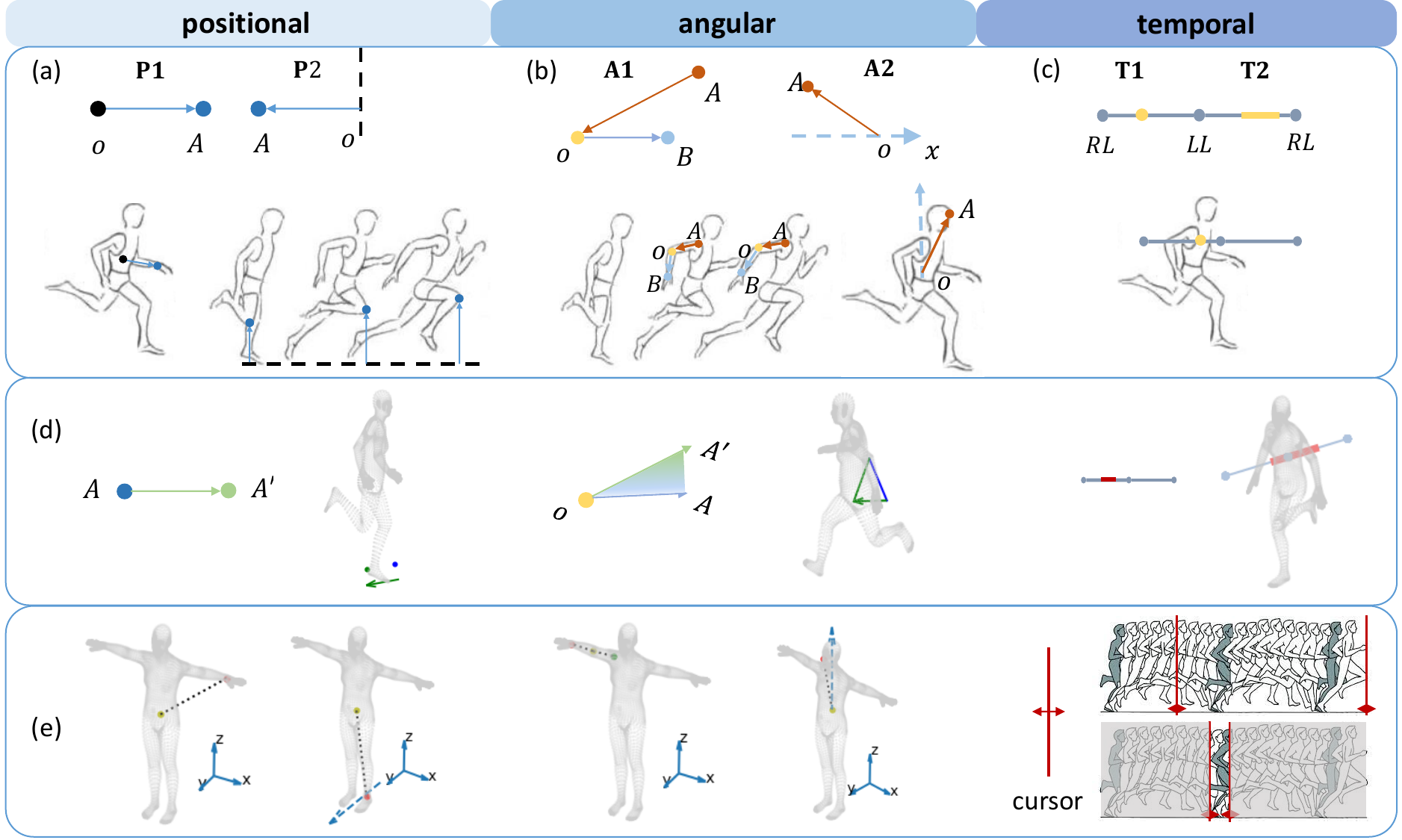}
  \vspace{-3mm}
  \caption{The three sets of mappings.
    The first mapping is the representation of positional (a), angular (b) and temporal (c) attributes. 
    The second mapping is the representation of their differences (d).
    The third mapping is the user operations required to define an attribute of each type (e).
  }
\label{fig:mappings}
\vspace{-1mm}
\end{figure*}

{This section introduces {the} data (pose attributes) we study.
Sports biomechanics generally includes kinematics (movements) and kinetics (forces)~\cite{bartlett:2014:IntroSpBio}.
Our system focuses on the former since movements are directly observable from videos, while forces require physical sensors (e.g., force plates) for measurements.
To support user-defined pose attributes, we classify {kinematic} human pose attributes from a common biomechanics perspective~\cite{bartlett:2014:IntroSpBio,novacheck:1998:BR,van:2021:BRS} into positional, angular, and temporal.}
This forms a design space of pose attributes, containing the representation of pose attributes ({\textbf{R3}}), the representation of their differences for showing feedback ({\textbf{R1}}), and user operations required to interactively define the pose attributes ({\textbf{R4}}).
In the following we introduce these three components of the design space in detail.

\para{Pose Attributes.} We first introduce the pose attributes as follows:

\begin{itemize}[leftmargin=*]
    \item \textbf{Positional attributes} (\autoref{fig:mappings}(a)) are defined as the relative distance between two points (classified as type \textbf{P1}), or the position of a point from a specific axis (\textbf{P2}).
For example, the trajectory of the wrist is its relative distance to the body center (\textbf{P1}).
Another example is the knee lift, which is a vertical distance from the knee joint to the body center (\textbf{P2}).

\item \textbf{Angular attributes} (\autoref{fig:mappings}(b)) are defined as either the angle formed by three endpoints (classified as type \textbf{A1}), or the orientation of a vector formed by two joints with respect to an axis (\textbf{A2}).
For example, the elbow angle (\textbf{A1}) is an angle formed by the shoulder, the elbow and the wrist joint.
The leaning of the upper body (\textbf{A2}) is the orientation of the vector pointing from the root joint to the neck joint w.r.t. the z-axis.

\item \textbf{Temporal attributes} are defined as either a single moment (\textbf{T1}) or a time range within a running cycle (\textbf{T2}). 
We use a temporal axis to show the temporal context.
The temporal axis (\autoref{fig:mappings}(c)) is a fixed full running cycle, with the three dots from left to right respectively corresponding to the states of right foot landing ($RL$), left foot landing ($LL$), and right foot landing for the next cycle.
The positioning of the human center on the temporal axis reflects the state of the current pose within the running cycle.
\end{itemize}

\para{Attributes Differences.} Pose attributes differences are mainly used for presenting feedback, i.e., from an incorrect configuration to a correct one.
We define a set of visuals for attribute differences (\autoref{fig:mappings}(d)), which are unified with the attribute representation.
Positional difference is shown by two points and an arrow pointing from the wrong position to the correct position.
Angular difference is shown by two vectors forming a wedge to show an angular difference.
Temporal difference is represented by a red marker segment on the temporal axis showing a temporal offset.
For example, the red segment along the forward temporal axis direction indicates the current event should appear later.

\para{User Operations.} In this section we introduce the user operations (\autoref{fig:mappings}(e)) for defining their own pose attributes under the three data attribute classes.
Specifically, the query editor in our user interface (\autoref{fig:interface}(e)) contains a 3D viewer presenting the 3D human body model (SMPL~\cite{Matthew:2015:SMPL}) in T-pose, radio buttons for specifying properties and two draggable cursors (red lines) on top of a running cycle diagram for specifying timings.
A user may either refer to the mesh or skeleton of the body model and directly mouse-click on the body model to select joints; our system will snap the mouse click to the nearest joint.

A user first selects the attribute type by selecting either the angle button or distance button for angular and positional attributes, respectively, or directly dragging the temporal cursors for a temporal attribute.
To edit a positional attribute, a user first specifies the joint to track, and then specifies the base point (\textbf{P1}).
When the user further selects an axis, only the component of the selected dimension will be recorded (\textbf{P2}).
To edit an angular attribute, a user either selects three endpoints in order on the body model (\textbf{A1}), or two points and one axis (\textbf{A2}).
To edit a temporal attribute, the user either moves one cursor to specify a moment (\textbf{T1}), or both cursors to specify a time range (\textbf{T2}).
Our system will record a phase or a phase range accordingly.
When the positional and angular attributes are associated with an event, the user also moves the temporal cursor to specify the timing.
Please refer to the demo video for the authoring process of attribute examples.

\subsection{Data Analysis Model}
\label{sec:method}

In this section we introduce the methods of the backend modules in \emph{PoseCoach} (\autoref{fig:architecture}): video processing, pose analysis, and feedback.

\subsubsection{Video Processing}

\begin{figure}[!ht]
  \centering
  \includegraphics[width=\linewidth]{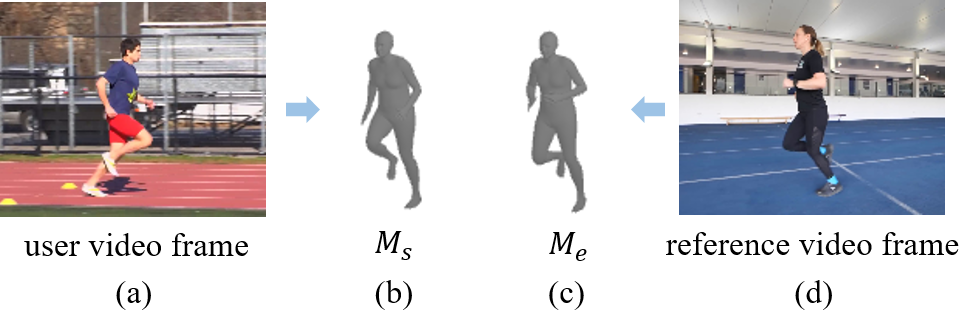}
  \vspace{-9mm}
  \caption{The postures in the user and the reference videos are aligned spatio-temporally.
  Each frame in the user video (a) is temporally aligned with a frame in the reference video (d) according to the similarity in posture joint rotations. The reconstructed poses from both frames (b-c) are unified to the same global orientation for comparison.
  }
\label{fig:alignment}
\vspace{-3mm}
\end{figure}

\para{3D Pose Reconstruction and Normalization.}
When the user and the reference videos are loaded into the system, the pose at each frame is retargeted onto a SMPL model.
{It is} achieved by {the following reconstruction and normalization steps}.
The poses in video frames are first reconstructed with the 3D SMPL model.
We adopt an existing pose reconstruction method in our implementation, TCMR~\cite{choi:2021:TCMR}, which achieves the state-of-the-art accuracy on challenging outdoor video datasets.
{Denote} the reconstructed poses as $M_s$ for the {user} video and $M_e$ for the {reference} video.
{$M_s$ and $M_e$ are normalized to the same body configurations (i.e., height and bone lengths); they are also normalized (\autoref{fig:alignment}) to a unified global orientation to factor out variations in viewpoints (\textbf{R2}).}

\para{Temporal Segmentation.}
The running pose sequences in both the user and reference videos are segmented by the key frames of foot landing and foot extension.
Since the action of running is periodical, we adopt the phase variable of human locomotion, as in~\cite{holden:2017:LMP}.
A full running cycle thus contains four key phases, in ``right foot landing'' ($phase=0$), ``right foot extension'' ($phase=0.25$), ``left foot landing'' ($phase=0.5$), and ``left foot extension'' ($phase=0.75$) order.
These four key phases are detected from the local extrema of the foot trajectories.

\subsubsection{Pose Analysis}

\para{Sequence Alignment.}
Given the detected key phases, the running pose sequences in the user and reference videos are first temporarily aligned at key phases, and then aligned at a finer level between each two key phases using the dynamic time warping (DTW) technique~\cite{berndt:1994:DTW}.
{The DTW algorithm finds the optimal alignment between the two sequences using the skeleton-agnostic joint rotations to measure human pose similarity {and thus handle} the variations in running speed (\textbf{R2}).}

\para{Attributes Retrieval.}
Each pose attribute is represented with a meta data tuple: $\left[name, type, J_A, J_o, J_B, axis, side, phase \right]$,
where $J_A, J_o, J_B$ are the joint IDs of the attribute endpoints in the body model (as shown in \autoref{fig:mappings});
$side$ is one of the ``left'', ``neutral'' and ``right'';
$axis$ and $phase$ are the related axis and timing of the attribute; they are left empty if not applicable.
For customized attributes, the meta tuple is formed from users' input from the query editor.
Our attribute retrieval program parses the meta tuple and outputs retrieved values from the videos, so that the customized attributes can be retrieved in the same way as the pre-defined attributes ({\textbf{R4}}).
The retrieved values are then used for comparison.

\para{Comparison.}
Since different attributes have different scales and units, we normalize the attribute values to the range $\left[0,1 \right]$.
Then the differences in the attribute values are computed as the relative errors between the attributes from the user video and those from the reference video.
We set a threshold of $25\%$ to select the significantly different attributes and visualize them to users as feedback.

\subsubsection{{Feedback to the User}}

\para{Animation-based Demonstration.}
The corrective suggestion from pose comparison is conveyed by animating a 3D human model ({\textbf{R3}}).
To make the demo easily understandable, the animation follows the design guideline as data-GIF~\cite{shu:2020:GIF}.
The animation contains two key frames corresponding to the wrong pose and the same pose with a specific body part in the position as the reference pose, respectively.
Specifically, we use the joint rotations to drive the model: for angular attributes, the intermediate frames are interpolated with the joint rotations of $J_o$; while for positional attributes, the animation is interpolated with the joint rotations of the parent joint of $J_o$ along the kinematics tree.
The 3D animations are augmented with visual markers to highlight differences ({\textbf{R1}}), as in~\autoref{fig:mappings}(b).

\para{Suggestive Viewpoint.}
Since the animation of corrective suggestion is in 3D, we would like to demonstrate it at the most informative viewpoint. While there are prior studies on the automatic selection of viewpoints for previewing a 3D mesh, the definition and criteria of the optimal viewpoints are often dependent on the purpose, such as to demonstrate region visual saliency~\cite{lee:2005:MSY}, to set man-made models in upright orientation~\cite{fu:2008:UOM}, and to incorporate modelers' creation processes~\cite{Chen:2014:HVA}.
Previous studies on optimal viewpoints for human poses mainly include reducing prediction uncertainty in estimating 3D pose~\cite{Kiciroglu:2020:AMC} and metrics defined over body part visibility~\cite{Kwon:2022:OCP}.
In \emph{PoseCoach}, since we would like to provide suggestions w.r.t. specific 3D local pose attributes, we develop a set of schemes to suggest viewpoints according to the geometry of the attributes.

\begin{figure}[!h]
  \centering
\includegraphics[width=0.95\linewidth]{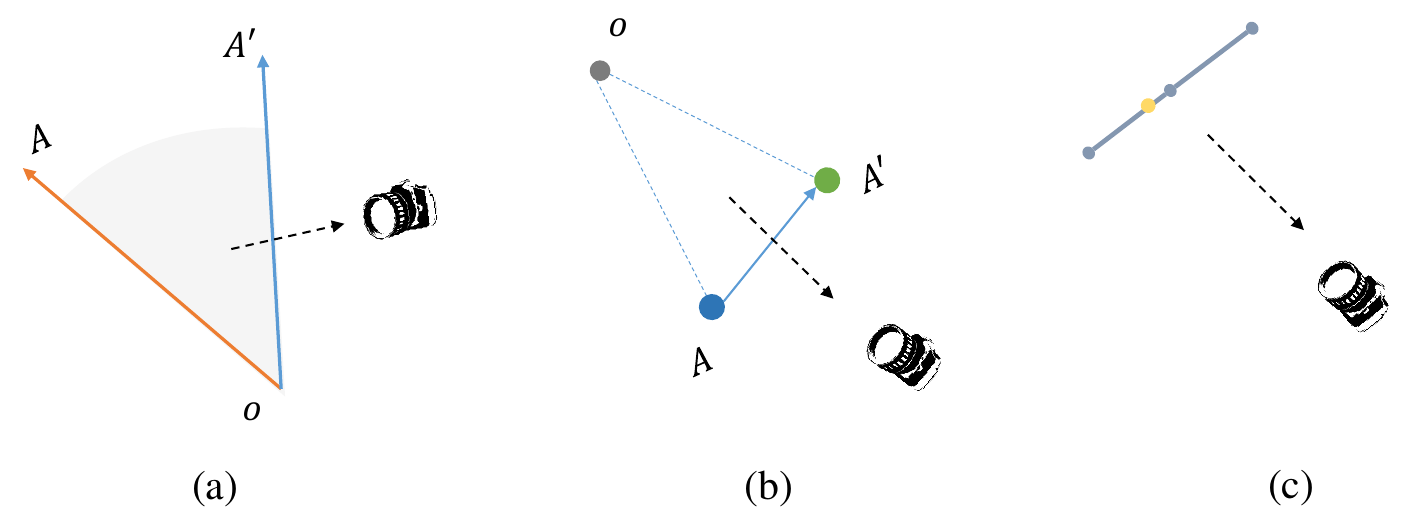}
\vspace{-3mm}
  \caption{Deciding viewpoints by the geometry of the 3D attributes.
    (a) For angular attributes the viewpoint is along the normal of the plane where the angle lies in, pointing outwards the body model; (b) for positional attributes the viewpoint is along the normal of the plane formed by the wrong position, the correct position, and the body center; (c) the temporal axis is within the Sagittal plane of human, and the viewpoint is perpendicular to this plane.
  }
\label{fig:cameraPos}
\vspace{-3mm}
\end{figure}

The main idea is to minimize the ambiguity in the attributes due to camera projection ({\textbf{R2}}), while preserving the human model as the spatial context.
Based on this goal, we make use of the normal vector formed by the 3D attributes to decide the orientation of the viewpoint (see \autoref{fig:cameraPos}).
We further use the side of the body to determine whether to revert a normal to its opposite direction.
For example, to present an attribute on the right side of the body, the camera should also be placed to the right facing the body model.
The up direction of the viewpoint is along the average of the two vectors.
We also determine whether to revert the up direction according to whether it keeps the human model heading upwards.
Even though we present the 3D animation in the suggested viewpoint, users can still manually change the viewpoint to explore the corrective suggestion ({\textbf{R4}}).

\section{Visual Design}
\label{sec:systemdesign}

In this section we first describe the predefined data (template pose attributes) in \emph{PoseCoach} in Sect.~\ref{ssec:attributes}.
Then in Sect.~\ref{ssec:overviews} we introduce the visual designs for showing pose attributes feedback, including glyphs and a timeline.

\subsection{{Template Attributes}}
\label{ssec:attributes}

\begin{figure}[!ht]
  \centering
  \vspace{-2mm}
\includegraphics[width=0.97\linewidth]{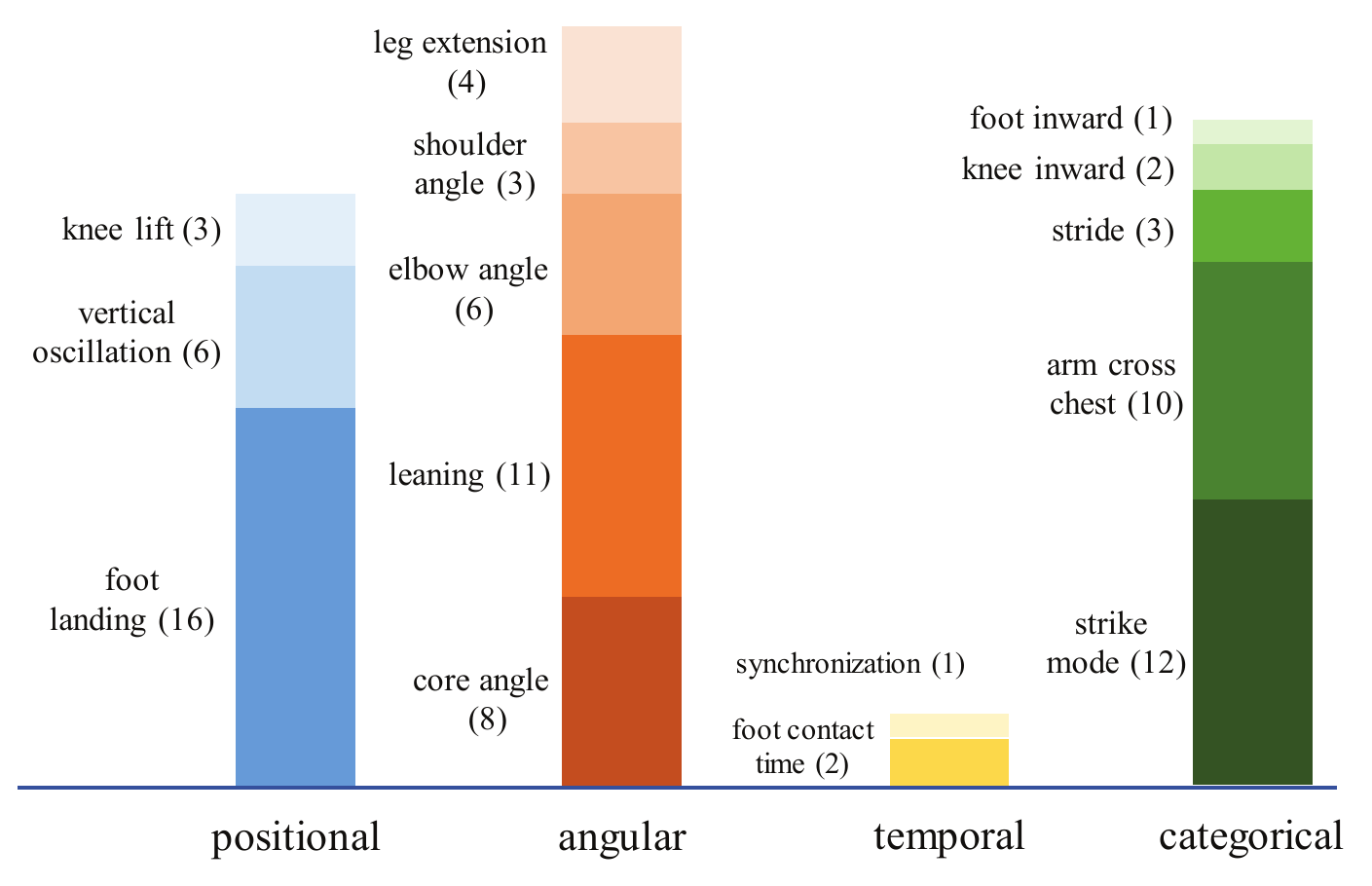}
  \vspace{-4mm}
  \caption{Common pose attributes for evaluating running poses summarized from our collected corpus.
  In the brackets are the numbers of their occurrences in the corpus.
  }
\label{fig:data}
\vspace{-4mm}
\end{figure}

{Since there are common data attributes for evaluating running pose correction, \emph{PoseCoach} embeds several pose attributes as templates to simplify user operations.}
To decide the template pose attributes, we collected a corpus of running pose tutorials by searching with key words ``running pose tutorials'', ``running pose corrections'', ``running techniques'', ``running form'', etc., from Google and YouTube.
The resulting corpus contains 55 items (including 37 videos and 18 articles). The pose attributes are summarized from the corpus into four types, as shown in \autoref{fig:data}.
{Different from the previously discussed three types (i.e., positional, angular, and temporal) in Sect.~\ref{ssec:mappings},} the fourth type ``categorical data'' are not computed from comparison with reference poses, but computed directly based on the other three types.
{For example, the pending of knee inward or outward (categorical data) is via the computation of knee angle (angular data).}
Thus we focus on the design for {the three types of data in Sect.~\ref{ssec:mappings}}, but support the visualization of the categorical data for commonly evaluated attributes in running.
We conducted another interview with \textbf{E4} to verify the coverage of these attributes in running pose evaluation in practice.

\subsection{Design of Attributes Overview}
\label{ssec:overviews}

In this section we discuss the design of the attribute overview for the problems reflected from the comparison.
The overview should show which attributes appear in question in the user video and their timings ({\textbf{R4}}).
We thus propose to use glyphs to represent attributes and a timeline tailored for running to organize them temporally.

\begin{figure}[!h]
\vspace{-2mm}
  \centering
\includegraphics[width=0.95\linewidth]{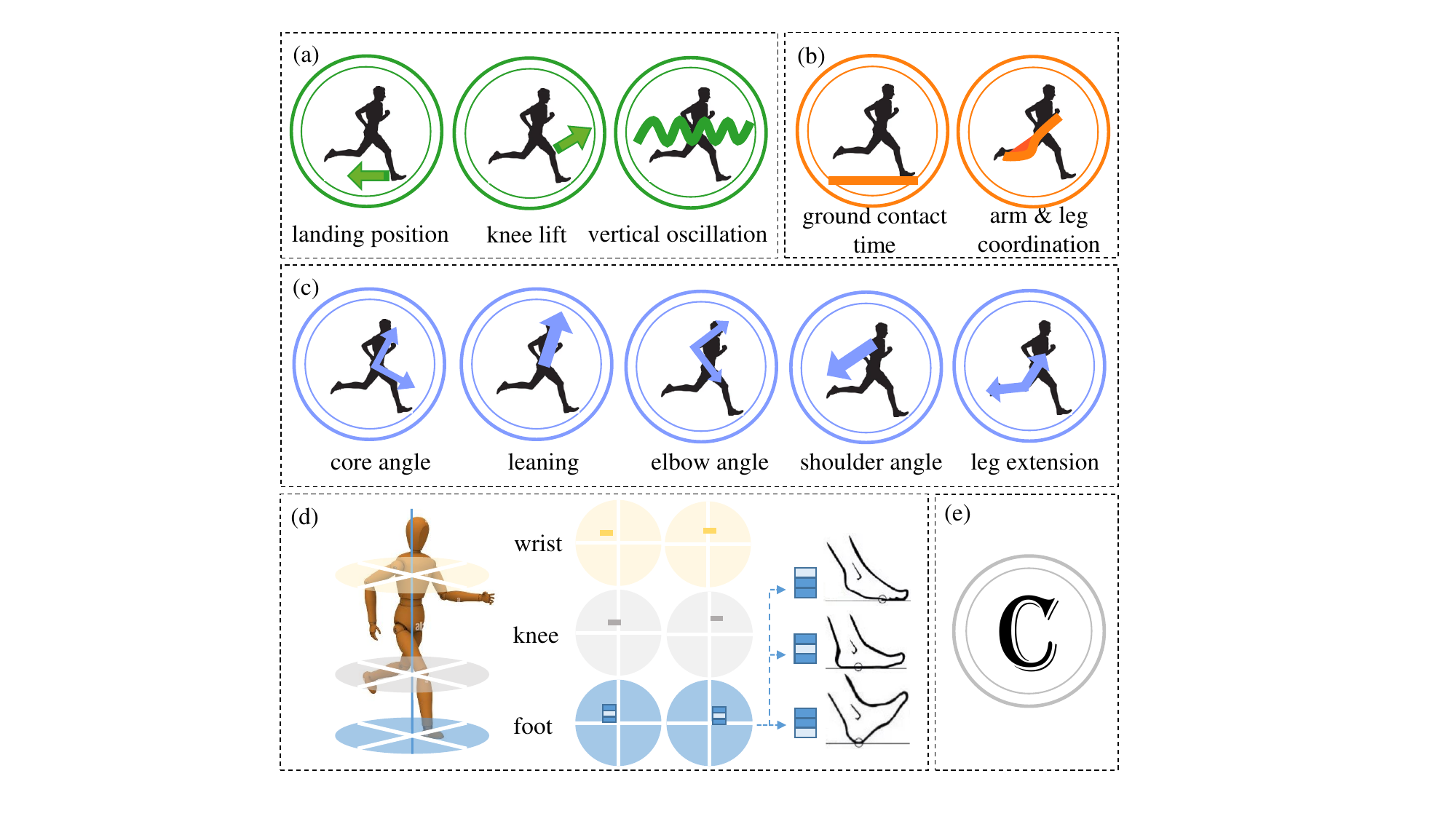}
\vspace{-2mm}
  \caption{The glyphs for representing attributes.
  For suggestion glyphs, the color of the circles encodes the attribute type: green for positional (a), orange for temporal (b), and blue for angular (c).
  Profile glyphs (d) use floating rectangles overlaid on static circles to show the positions of wrists, knees and feet in three transverse planes.
  The rectangle for the foot joint also encodes the fore-foot, middle-foot, or rear-foot strike mode.
  We design another glyph (e) to represent a customized attribute on the timeline. Note that we have changed the color after the user study based on reviewers' feedback, and we believe this change should not have a significant impact on our user study results since we did not specifically ask about the color scheme nor received any specific comments about colors from the study participants.
  }
\label{fig:icons}
\vspace{-1mm}
\end{figure}

\para{Glyphs.}
We designed two types of glyphs for the four classes of the attributes, namely suggestion glyphs and profile glyphs.
{Suggestion glyphs are icons for each of the three classes of attributes in Sect.~\ref{ssec:mappings}, i.e., positional, angular and temporal attributes,} whose values are continuous variables and are compared with those in the reference videos.
As shown in~\autoref{fig:icons}(a-c), {we augment a symbol of a running pose with markers to represent the semantic meaning of a pose attribute, and use the categorical color scheme to distinguish the three types of data.} 

The profile glyphs are used to represent categorical attributes, {which are not derived from comparisons with the reference, 
but from the relative positions of the joints in the categorical attributes (i.e., wrists, knees, and feet) relative to the body center.}
We adopt the idea from the dance notations~\cite{mirzabekiantz:2016:BMN} to discretize complex human movements at reference planes (sagittal,
frontal, and horizontal).
As shown in~\autoref{fig:icons}(d), the profile view contains three transverse planes showing a summary of the projections of {the} three joints of the left and right sides in their respective key timings in a stride.
Specifically, to indicate whether the wrists cross the body's middle line in front of the chest, we use the positions of the rectangles to show the closest horizontal displacements of the wrists relative to the body center during running.
Similarly, whether knee and foot cross the middle line can be visualized by their relative positions to the body center at landing.
In the transverse planes for the feet, besides showing the landing position relative to the body center, the triple stacked rectangles further indicate the strike mode (fore-foot, mid-foot or rear-foot strike) of each foot by highlighting one of the rectangles representing the corresponding position.

\para{Timeline.}
A characteristic of a running pose attribute sequence is that it is temporally periodical, and each period can be divided into a right-phase and a left-phase.
Based on this characteristic, we propose to design a timeline that transforms the temporal space into a running event space{, which is robust to the variations in video lengths (\textbf{R2})}.
As shown in \autoref{fig:teaser}(c), the horizontal axis is a complete running cycle, and the vertical axes correspond to the attributes of the left side of the body, right side of the body, and middle, respectively.
All the data attributes are summarized among cycles to be shown on the timeline.
Our system will automatically select significant errors, with the sizes of the glyphs proportional to the significance of the errors of a particular type.

\begin{figure}[!h]
  \centering
  \includegraphics[width=\linewidth]{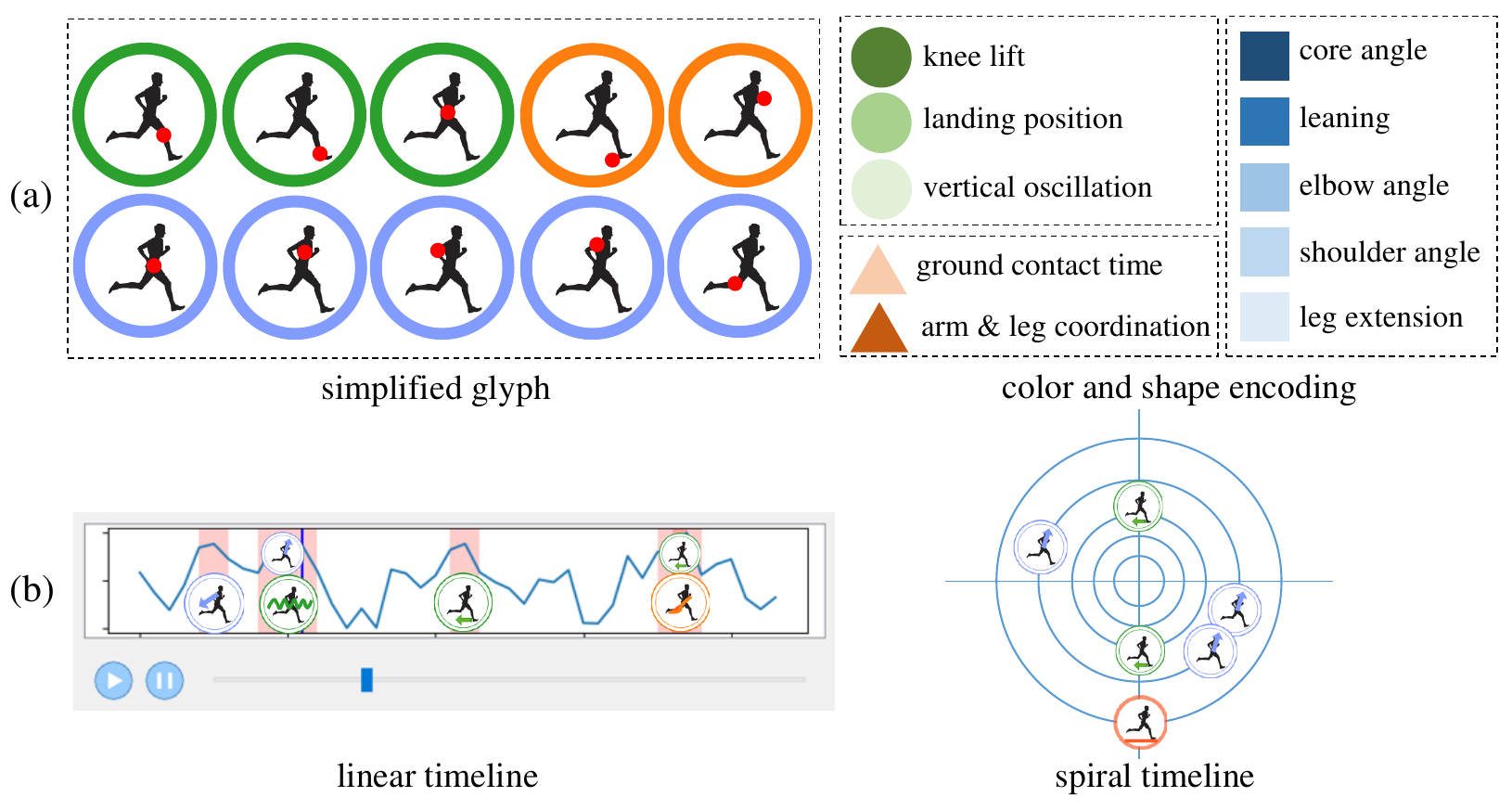}
  \vspace{-7mm}
  \caption{Alternative designs for: (a) icons and (b) timeline.}
\label{fig:alternatives}
\vspace{-4mm}
\end{figure}

{We examined the design of {the} suggestion glyphs and timeline against their alternatives (see \autoref{fig:alternatives}).}
{The alternatives for glyph design} include a set of simplified icons highlighting the body parts in question and color and shape encoding.
{The glyphs augmented with markers are more intuitive and easier to remember than these alternatives.}
For timeline design, the alternatives are an ordinary linear timeline (\autoref{fig:alternatives}(b) left) and a spiral timeline (\autoref{fig:alternatives}(b) right). 
In the linear timeline, a video is not segmented into running cycles, and the background shows the plot of a selected attribute. 
In the spiral timeline, each ring corresponds to a running gait cycle, and the glyphs display all occurrences of attribute differences.
{The aggregated timeline provides a better summary of problems than {the} linear and spiral {timelines} while retaining their timings with respect to the running gait cycle.}

\section{User Evaluations}
\label{sec:result}

In this section, we show the results of a user study {to evaluate} the visualizations of posture correction feedback in \emph{PoseCoach} and the baseline methods (Sect.~\ref{sec:userstudy}) for novices, expert interviews (Sect.~\ref{sec:Einterviews}) to evaluate the overall effectiveness of the system, and a validity test of our system conducted with amateur runners (Sect.~\ref{sec:realdata}).

\subsection{In-lab Comparative Study}
\label{sec:userstudy}

The main purpose of the user study is to evaluate the improvement of {\em PoseCoach} in {promoting novices' perception of running} pose differences over existing methods (see \textbf{Baselines}).
It also evaluates the effectiveness of other components (e.g., viewpoint navigation and summarization of feedback) in assisting novices' perceptions of running pose improvements.

\para{Apparatus.} {We implemented} \emph{PoseCoach} with Python 3.6 on a PC running Win10 (Intel x64 i5 CPU@3.00GHz, 8.00GB RAM).
The user interface was implemented with the PyQt5 framework.
Due to the current local COVID-19 regulation, the user study was conducted via {\em Zoom} with remote screen control.

\begin{figure}[!ht]
\vspace{-2mm}
  \centering
  \includegraphics[width=0.98\linewidth]{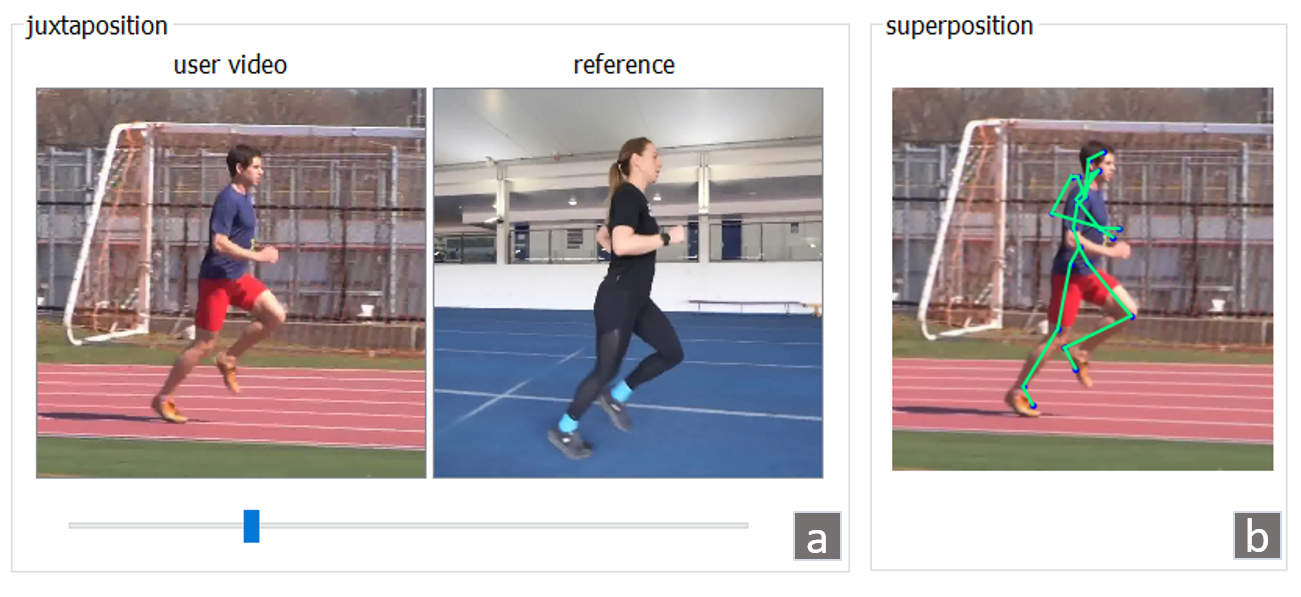}
  \vspace{-3mm}
  \caption{The UI for the baseline methods:
  (a) juxtaposition: compare two synchronized frames side-by-side;
  (b) superposition: overlay the transformed pose in the reference video frame onto the user video frame.
  }
\label{fig:baselines}
\vspace{-2mm}
\end{figure}

\para{Baselines.}
The baseline methods are visualizations of pose differences via juxtaposition and superposition, as shown in~\autoref{fig:baselines}.
We implement the baselines as follows.
For juxtaposition, we used the setup in~\cite{CoachesEye:2011:CE} and put two running poses side-by-side.
To facilitate the preview, the two poses are cropped with subjects' bounding boxes in videos, and the two videos are temporally synchronized using joint rotations.
For superposition, we adopted the method in~\cite{clarke:2020:ReactV}. 
Since~\cite{clarke:2020:ReactV} is based on Kinect, we transformed the 3D pose in a temporally correspondent reference video frame and aligned it to the pose in the user video frame at the body center, such that the temporally synchronized reference pose is overlaid on the user video frame for comparison.

\para{Participants.}
12 members from a local university were invited to participate in the user study (a1$\sim$a12, aged 23$\sim$32, 3 female).
Except for a1 and a7, all the other participants practise running more than once a week, but do not have access to professional coaches.
a12 stated that he was once curious about the correctness of his running poses and searched for mobile apps providing running pose checking functions but could not find a suitable one.
a2 focused on foot landing during running to avoid injuries; a6 used body senses after running as feedback.
a3, a10 and a11 said that they used mirrors during fitness workout, but obtained no feedback on pose correctness during running.

\para{Task.}
To ensure that the participants {were} under the same condition to facilitate {fair} comparisons, we prepared 9 user videos (V1$\sim$V9) covering all of the ten pre-defined {attributes}.
They were collected from running tutorial videos such that the ground-truth of the mistakes in running poses was known from the coaches' comments in the videos, such as foot landing in front of the body (the braking position) and insufficient knee lift.
The difficulty level of the videos was controlled by containing only one main problem.
The general task for the participants was to explore the corrective feedback from videos using either \emph{PoseCoach} or the baseline methods in a think-aloud manner, and complete a questionnaire afterwards.
The user study contained three sessions: two sessions using our system with and without the suggestive viewpoints, and one session using the baseline methods.
The order of the three sessions was counter-balanced, and the order of the nine videos was randomized among the three sessions (three videos for each session).
During training, we first gave a detailed tutorial on the operations of \emph{PoseCoach} as well as the baseline system.
The participants then tried freely to get familiar with both systems.

In the session using \emph{PoseCoach} without suggestive viewpoints (denoted as {``PoseCoach-w/o''}), we disabled the suggestive viewpoint function, and the participants would need to manually navigate the viewpoints to preview the 3D animations.
The system recorded the participants' navigation activities in the suggestion preview window, parameterized by viewpoint azimuth and elevation, and the duration of each viewpoint.
In another session using \emph{PoseCoach} (denoted as ``PoseCoach''), the suggestive viewpoint function was enabled; the participants could also manually navigate, and their navigation activities were also recorded.
In the session using the baseline methods (denoted as ``Baseline''), the participants explored the corrective feedback by comparing running poses in videos in either juxtaposition or superposition visualization.

After the sessions, the participants completed a designed questionnaire (\autoref{table:questionnaire}) in a 7-point Likert Scale (1 is Strongly Disagree and 7 is Strongly Agree), and a standard System Usability Scale (SUS)~\cite{brooke:1996:SUS}.
The user study with each participant took about 90 minutes.

\begin{table}[!h]
\begin{tabular}{p{0.5 cm}|p{7.2 cm}}
\toprule
Q1 & The feedback of posture correction is easy to access. \\ 
Q2 & The demonstrations of pose differences are easy to understand. \\ 
Q3 & The visual designs are intuitive. \\
Q4 & The feedback reflects the problems in user videos. \\
Q5 & The feedback is helpful in improving running postures.  \\ 
\midrule
Q6-Q9 & Demonstrations with animation, normalized poses, summary of mistakes, suggested viewpoints are helpful for understanding suggestions. \\
Q10 & I'm more satisfied with \emph{PoseCoach} than only browsing videos and overlaid poses. \\
\bottomrule 
\end{tabular}
\vspace{2mm}
\caption{The questionnaire used for the user study:
Q1$\sim$Q5 evaluate the effectiveness of \emph{PoseCoach} in providing corrective feedback on running poses.
Q6$\sim$Q10 evaluate the advantages of design components in \emph{PoseCoach} over the baseline methods.
}
\label{table:questionnaire}
\end{table}

\begin{figure}[!ht]
  \centering
  \vspace{-7mm}
\includegraphics[width=\linewidth]{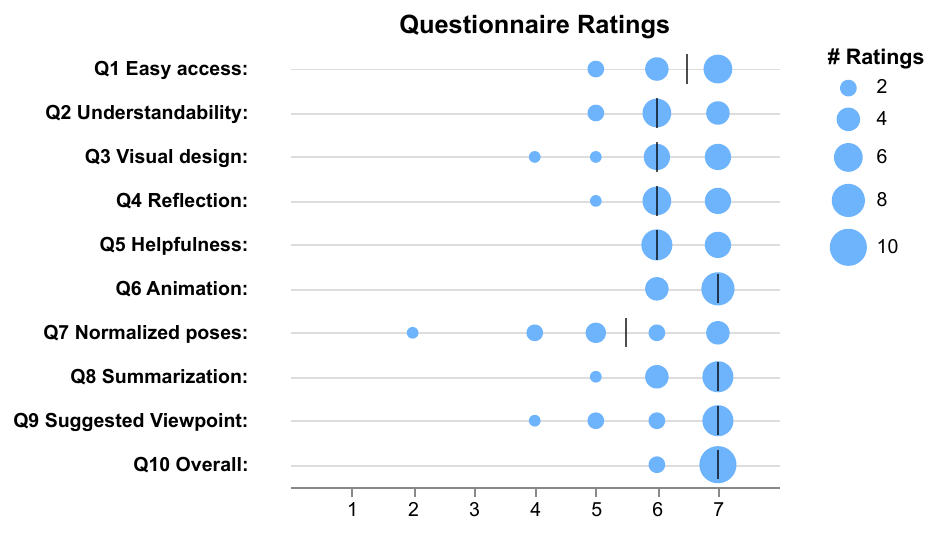}
\vspace{-7mm}
  \caption{Summaries of the participants' responses to the questionnaire in \autoref{table:questionnaire}.
  The black vertical bars represent the median of the ratings for each question.
  }
\label{fig:likert}
\end{figure}

\begin{figure}[!ht]
  \centering
  \vspace{-3mm}
  \includegraphics[width=\linewidth]{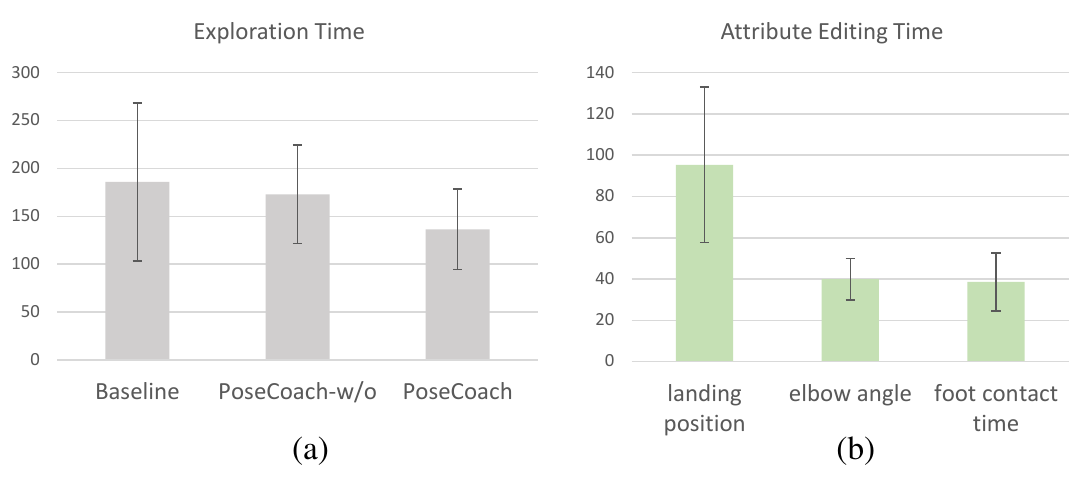}
  \vspace{-8mm}
  \caption{Statistics on time in the user study. (unit: s).
  (a) Average exploration time for using the baseline methods, \emph{PoseCoach} without suggestive viewpoint, and \emph{PoseCoach} with suggestive viewpoint.
  (b) Average editing time for the three selected attributes using the query editor.
  }
\label{fig:times}
\end{figure}

\para{Effectiveness of Pose Difference Visualization.}
We first investigate the effectiveness of \emph{PoseCoach} in presenting feedback compared with the baseline system.
Q10 explicitly asked the comparison between \emph{PoseCoach} and the baseline methods, where 10 out of 12 participants strongly agreed that \emph{PoseCoach} was more effective in conveying feedback than the baselines.
We recorded the time required to explore the running pose problem(s) in each video, as shown in \autoref{fig:times}(a).
Paired t-tests on {the} exploration time required for each video among sessions showed that using \emph{PoseCoach} with the suggestive viewpoint significantly requires less time to obtain the desired feedback than both {\emph{PoseCoach}} without the suggestive viewpoint ($p=0.012$) and the baseline system ($p=0.019$).
However, there is no significance on exploration time between sessions ``PoseCoach-w/o'' and ``Baseline'' ($p=0.519$).

We evaluated the accuracy via the successful rate of the participants' discovered mistakes matched the ground-truth mistakes as commented by the coaches in videos.
In sessions ``PoseCoach-w/o'' and ``PoseCoach'' the successful rate was $100\%$. 
In other words, all the participants could figure out the problem(s) in the running poses with the visualization provided by \emph{PoseCoach}.
In contrast, the successful rate was $77.8\%$ in session ``Baseline''.
From the participants' think-aloud in session ``Baseline'', they often referred to the superposition visualization more than the juxtaposition visualization, especially when the subjects in the user and reference videos are running in different directions.
For superposition in the baseline system, a6 and a8 said that they would refer to the lower limbs more often than upper limbs, since upper limbs were often occluded and misaligned due to differences in limb lengths.

\para{Effectiveness of System Components.}
We then investigate the influence of specific design components on users' perception of feedback on running pose correction.
Q6 asked the participants to rate the key component in \emph{PoseCoach}, which visualizes pose differences via animations of local body parts on a human model.
8 out of 12 participants strongly agreed that such visualization was helpful for understanding, and the other four chose agreed.
The component that received the most disagreement is the preview of normalized poses from the user and reference videos shown in juxtaposition (\autoref{fig:teaser}(a) middle).
Since their orientations are often different from those in the original videos, the participants stated that referring to them increased the cognitive load by having to imagine the transformation to understand.
Thus even though normalized poses are crucial to computing pose differences, they do not necessarily contribute to users' visual comparison.
During the participants' think-aloud in sessions ``PoseCoach-w/o'' and ``PoseCoach'', they often directly moved on to check the glyphs on the timeline after loading both videos.
After watching the animation, they sometimes checked the user video frame to verify the problem.
At first they sometimes also referred to the reference video frame to verify the animation, but many of them skipped the reference video frames later because they found the corrective feedback illustrated by the animation was trust-worthy.

We also evaluated the usefulness of the design component of suggestive viewpoint.
We would like to figure out the following two questions: (1) do users find previewing the animations of pose correction under a certain viewpoint yields better perception?
(2) If yes, do our suggestive viewpoints match the preferred viewpoints selected by users?
We thus analyze the usage of viewpoint selection during the user study.
In session ``PoseCoach-w/o'', the average number of times the participants manually changed the viewpoint was 7.36 times per video, compared with 2.05 times per video in session ``PoseCoach''.
A paired t-test on the numbers of manual navigation between sessions ``PoseCoach-w/o'' and ``PoseCoach'' shows that enabling the suggestive viewpoint function significantly reduces users' manual navigation ($p=0.00059$).
To answer question (2), we further analyze the relevance of the participants' manually-selected viewpoints with the suggested viewpoints computed by our system in session ``PoseCoach-w/o''.
We analyzed previewing viewpoints that lasted more than one second and considered those with a duration less than one second as the navigation process.
The average errors of azimuth and elevation relative to $360^\circ$ were $3.19\%$ and $4.52\%$, respectively, indicating a good match between our suggestive viewpoints and preferred viewpoints by the participants.

In the rating of the usefulness of suggestive viewpoint, seven participants chose ``strongly agree'', and four of them explicitly stated during exploration that this function was very convenient.
a2 in session S1 asked whether the suggestive viewpoint function could be enabled, because she found this function especially useful when she was comparing the magnitudes of corrections on foot landing position.
a4 found the suggestive viewpoint more useful in observing upper limbs because they often suffer from heavier occlusions by the body torso than lower limbs.
Interestingly, a12 rated ``Neutral'' in Q9.
He explained that since he studied exoskeleton robotics, he was more used to imagining the attributes using the sagittal, coronal and transverse planes as reference, rather than using the human body as a spatial context.
Since \emph{PoseCoach} targets at novice users without human movement analysis background, and most participants found the suggestive viewpoint function convenient, it can serve as a helpful option in \emph{PoseCoach}.

\para{System Usability.}
In the training session, all the participants could get familiar with \emph{PoseCoach} within 5 minutes by completing a pipeline of operations, including loading videos, previewing frames and poses, and navigating on the timeline to preview animations of suggestions.
The SUS score for all the ten questions in the SUS questionnaire was 83.125 on average (SD: 10.56), out of a scale of
100, indicating the good usability of \emph{PoseCoach}.

In post-study interviews with the participants, they commented favorably towards \emph{PoseCoach}.
For example, a3: ``\emph{Besides clarity, the summarization in \emph{PoseCoach} helps me form a better impression of frequent mistakes.}''
The participants also commented on the potential generalization of \emph{PoseCoach} in other scenarios.
Specifically, a11: ``\emph{This tool is solving a very practical problem. I can see how it is useful in running and can imagine it generalizes to many other sports.}''
a12 (from exoskeleton robotics background): ``\emph{... current rehabilitation training often relies on wearable sensors to detect patients' biomechanics, such as joint angular velocities and accelerations.
Such a video-based tool is promising in providing a non-invasive means to analyze patients' movements.}''

\para{Evaluation of Query Editor.}
From the user study we also evaluate the easiness of use of the query editor, specifically, how efficiently and accurately users can edit a pose data attribute.
There is no baseline method for this task.
We chose three frequently used data attributes from each of the classes in the pre-defined attributes, and asked the participants to edit the attributes using the query editor in our interface.
The three attributes were: ``foot landing position'' (\textbf{P2}), ``elbow angle'' (\textbf{A1}) and ``foot contact time'' (\textbf{T2}).
They covered all the operations on the query editor.
The participants were given user running video clips as references.
As shown in \autoref{fig:times}(b), the average editing time for the three attributes were 95.36s ($SD=37.71$), 39.91s ($SD=10.11$) and 38.64s ($SD=14.03$).
On average the editing of the foot landing position took the longest time, since it required the most operations covering all the components on the query editor.
The successful rates that the participants can implement the same attribute as our pre-defined was $83.3\%$, $100\%$, and $91.7\%$, respectively.
In the failure cases, a3 failed the temporal attribute, because he misunderstood the question and labeled the time between two consecutive foot landings instead.
a4 and a10 both correctly annotated the positional attribute on the human model, but forgot to associate with the timing for foot landing by dragging the timeline cursor.
Through this experiment we verified that novice users could easily understand and implement the representative attributes with minimal training.
Even though for most amateur runners the pre-defined attributes would suffice, they can annotate their interested attributes via the query editor with reasonable efforts.

\subsection{Expert Interviews}
\label{sec:Einterviews}

We conducted expert interviews to evaluate the overall usefulness of our system in helping amateur runners correct running poses.
Two experts with running backgrounds were invited: one was a licensed running coach (\textbf{E5}); the other was a professional marathon runner (\textbf{E6}).
The two interview sessions were conducted separately, and each session lasted 50 minutes.
During the interviews we provided a detailed introduction of functions in \emph{PoseCoach} with three demonstrations of usage scenarios, and then invited them to try the system freely.

Both experts strongly agreed that \emph{PoseCoach} would benefit a lot of runners.
\textbf{E5}: ``\emph{Not only beginners, but experienced runners are also often bothered by the problems of running form. I can expect this tool will serve a lot of runners.}''
They also appreciated that the design rationale of \emph{PoseCoach} is very reasonable for practical usage.
\textbf{E5} said that coaching is a highly personalized process; and thus a significant advantage of the design of \emph{PoseCoach} is that it does not directly classify a runner as right or wrong, but retains the flexibility to compare with various running poses to show the differences.
\textbf{E5} finds \emph{PoseCoach} especially useful for novices to iteratively adjust to different references to find their most suitable poses.
\textbf{E6} commented that the design of \emph{PoseCoach} is similar to the idea of the ``champion model'' for elite athletes, such as
Bingtian SU, who was trained by shortening the gaps (on both poses and capabilities) with elite exemplars.
This comment is consistent with \textbf{E3}'s advice in the formative study.

We also invited experts to comment on the positioning of \emph{PoseCoach} in training in real life.
\textbf{E5}: ``\emph{It is suitable for the majority of ordinary runners.
But for severely over-weight people, asking them to resemble the running of ordinary people might cause injury instead of reducing it; they should seek for professional advice instead.}''

\subsection{Validity Test}
\label{sec:realdata}

To evaluate \emph{PoseCoach} in practical usage, we invited five amateur runners (a13$\sim$a17, 2 females, aged 29$\sim$55) to analyze their running videos with our system.
% age (55, 52, 33, 29, 30)
These participants practiced running at least once a week (a13 and a14 had experience attending marathons), but none received professional training on running form.
In this user study, we asked them to collect their running videos and reference videos and describe what problems in their running forms they could identify {first without and then with \emph{PoseCoach}}.

We did not confine their ways of recording their videos. Thus they recorded their running process using various approaches: three participants had {another person record at a fixed position,} a15 used a tripod, and a16 had {another person} running after him while taking the video.
In choosing reference videos, a14 chose domestic professional marathon runners as references and used their training videos he searched and observed before. He said that nationality and gender {were} his major considerations because of the similarity in muscle strength.
Other participants chose references from famous runners they followed on {online video platforms (e.g., YouTube)}.
On average, selecting a reference video took about 8 minutes.

When asked to state the problems by observing the raw videos (and references), a15$\sim$a17 were able to describe a high-level problem, such as clumsy and rigid, while a13 and a14 stated that their running forms were perfect.
With \emph{PoseCoach}, those high-level {observations} are translated into concrete details.
For example, a17 found his upper body was unnatural as compared with the reference video, and the comparison suggested a larger leaning. {a14 compared his running form with three reference videos containing different subjects, and the results were consistent and showed only a slight difference in the knee bending angle.
a14: ``\emph{I have been carefully imitating these reference videos before, but seldom noticed this difference. I will pay more attention to the knee bending in future training.}''}

Besides the common attributes, a14 and a17 used the query editor to define new measurements for comparison.
Specifically, a14 edited the displacement of the elbow and back foot since these are the two attributes he was especially interested in when observing the reference videos.
a17 edited the horizontal displacement of the upper body because he thought his upper body had a noticeable difference with {the} reference.
a13 was interested in a prevailing running form, {the} Pose Method~\cite{romanov:2002:PMR}, and {used} the query editor to analyze her running poses w.r.t. this theory.
We noticed a limitation when a17 would like to analyze the {wiggling} magnitude of {the} upper body.
Currently, he edited the horizontal distance of shoulder movement for comparison.
A future direction is to design interactions to further support such a summary or statistics of pose attributes.

Finally, we invited the running coach \textbf{E5} to comment on the participants' running form{s} by analyzing the collected videos in the same way as his usual coaching process.
For each video, he pointed out a few key problems (the most for a13{, with three identified} problems), such as the ``braking posture'' when landing and large strides.
The problems \textbf{E5} mentioned were all reflected by the analysis results from the template attributes in \emph{PoseCoach}, while \emph{PoseCoach} provided more minor suggestions.
\textbf{E5} stated that these suggestions were not necessarily suitable for the participants, but in general, they would help reduce injuries.

\section{Discussions and Limitations}

\emph{PoseCoach} is designed to provide amateur runners {with} an approachable way to analyze their running forms in videos by comparing {pose attributes} with references. 
We learned a few key aspects from designing the system and limitations that require further exploration.

\subsection{Lessons Learned}

\para{Perspective Shortening in Videos.}
Video has the innate advantage of the ease of deployment, especially for in-the-wild sports, in which using physical sensors is impractical, making {videos} excellent media for analyzing movements.
Through the user study, we have confirmed the effectiveness of using reconstructed human poses in videos to provide running feedback to novices, as well as the importance of introducing viewpoints other than a video's original viewpoint.
Beyond human pose attributes, sports data often depend highly on 3D physical context, such as badminton shuttle trajectory~\cite{chu:2021:TIVEE}.
Using computer vision techniques to reconstruct 3D spatial-temporal processes from videos is likely to bring new insights through fusing the visualizations of the same process from multiple viewpoints.
By addressing the perspective-shortening issue, such as via retrieving 3D attributes and finding case-specific optimal viewpoints, the advantage of video-based analysis will be further enhanced.

\para{Positioning of \emph{PoseCoach}.}
The most frequent information we were reminded of from the expert interview was that there is no absolute correct running form.
As such, the sports pose analytics tools should be positioned as a reference that provides suggestions instead of professional advice for users.
According to \textbf{E3}, such an optimal pose variation among individuals also applies to other sports, such as swimming and tennis.
Running is comprehensive with many other factors than postures, where the overall goal is to promote the running economy~\cite{williams:1987:RE} (i.e., volume of oxygen consumed).
Thus as suggested by \textbf{E5}, our system is best used for helping different runners find their optimal ways of running according to the overall subjective comfort.

\para{Informative Design Components.}
As pointed out by \textbf{E3}, what athletes care about is the suggestions for improvement.
This is consistent with the participants' exploration processes in our user study.
When there is a direct suggestion (the animation), the participants did not need to frequently refer to the reconstructed 3D poses.
For novice users, the practical suggestions are more informative than the intermediate representations (e.g., normalized 3D poses).
Thus the design of visual analytics systems for novices should highlight the conclusions, instead of providing a fancy dashboard{, which would overwhelm} novices with abundant information.

\subsection{Scalability and Generalizability}

The most time-consuming component in our data analysis model is the deep learning-based human pose reconstruction (e.g., TCMR~\cite{choi:2021:TCMR} ran at about one fps on CPU), while the processing time of other components is negligible.
Since a running gait cycle normally lasts only a few seconds, users may input a short video (within 1 minute) to be processed within a reasonable time and obtain a summary of a few gait cycles.
Typically the running form of a video subject does not change in short videos.
\emph{PoseCoach} thus summarises the running form using the profile of an averaged gait cycle.
However, a video subject's running form might change within a long video, e.g., due to fatigue in marathon, and speed changes in Fartlek.
Temporal segmentation~\cite{aristidou:2018:DMS} for segment-wise summarizations will be required to handle different running forms within a video.

% generalize
This paper focuses on adults in moderate-speed running (jogging) since this is the most common type and demography for running exercises.
The current system is compatible with other running styles, such as sprint and up-hill run, using target exemplar videos as a reference.
Since comparisons of human poses are common in sports training, our overall design strategies of positional, angular, and temporal attributes can be adapted for other sports.
Specifically, positional and angular data are directly generalizable since they are derived from general human biomechanics.
Temporal attributes, as well as visual designs and pre-defined attributes, will need to be redesigned according to sport-specific domain knowledge, mainly key poses associated with key events, such as wrist angles for racket sports during striking.

\subsection{Limitations}

There are prior works~\cite{hanley:2018:DMV,reinking:2018:RTD,pipkin:2016:RVA} that validated the reliability of video-based biomechanics gait analysis as compared with MoCap or human raters in indoor controlled environments.
However, since we focus on in-the-wild running, it is impractical to apply motion capture to register the analysis results from our system and MoCap.
{Even though we adopt the state-of-the-art pose reconstruction method TCMR~\cite{choi:2021:TCMR},}
the reconstruction sometimes fails and affects the analysis at depth ambiguity (e.g., the bending of legs when recording from the back) and occlusions (e.g., one leg overlays the other when recording from the side view).
Currently the failures in pose reconstructions can be manually fixed using another interactive tool designed by us.
We believe that with the advancing monocular pose reconstruction techniques in computer vision, 
video-based sport pose analysis will be driven by a better backbone in the future.
Besides, in the evaluations, we mainly focus on assessing our core contribution, i.e., customizable pose analysis and visualization of pose differences.
The visual designs alone deserve a more comprehensive evaluation~\cite{Semeraro:2022:VIP} to gain insights on how novices from different countries and levels can best perceive feedback.
We will leave this as future work.

\section{Conclusion and Future Work}
\label{sec:conclusion}

We have presented a novel system, {\em PoseCoach}, for assisting amateur runners in improving their running forms.
We designed the system based on the design requirements formed from the literature research and expert interviews.
{\em PoseCoach} embeds common running pose attributes based on a collected corpus, and also provides an interface for users to customize attributes.
{\em PoseCoach} analyzes the poses from a user video and a reference video in 3D, and visualizes the pose differences via 3D animations on a human body model.
Our user study showed that demonstrating pose corrective feedback via 3D animations is more effective than displaying frames side-by-side or overlaying the correct poses onto the user video frames, {and our query editor is easy to use for novices.}

In future work, we would like to extend \emph{PoseCoach} with more {physiology}-oriented animations and analysis.
In the current setting the running pose attributes are analyzed and visualized independently.
But there are certain correlations among the attributes, e.g., a higher knee lift might yield a larger stride.
A potential improvement is to incorporate human body harmonics~\cite{kuhtz:2012:AUL,hinrichs:1990:WBM} to further summarize the problematic attributes.
Besides, currently {\em PoseCoach} focuses on the kinematics measurements (e.g., angles and positions).
However, more professional analysis~\cite{wei:2021:DPA} would require kinetics measurements, such as ground reaction force (braking force)~\cite{zecha:2018:SJF} and muscle elastic energy~\cite{Seth:2018:OS}.
Since the measure of kinetics parameters is currently limited to biomechanics laboratories, developing methods that recover the kinetics from videos would increase accessibility to many fields, including but not limited to sports posture analysis.

\section*{Acknowledgement}

We thank the anonymous reviewers for the valuable feedback and the user study participants for their great help. 
This research was supported by a grant from the Research Grants Council of HKSAR (Project No. HKUST16206722).

\bibliographystyle{IEEEtran}
\bibliography{ref}

\begin{thebibliography}{10}
\providecommand{\url}[1]{#1}
\csname url@samestyle\endcsname
\providecommand{\newblock}{\relax}
\providecommand{\bibinfo}[2]{#2}
\providecommand{\BIBentrySTDinterwordspacing}{\spaceskip=0pt\relax}
\providecommand{\BIBentryALTinterwordstretchfactor}{4}
\providecommand{\BIBentryALTinterwordspacing}{\spaceskip=\fontdimen2\font plus
\BIBentryALTinterwordstretchfactor\fontdimen3\font minus
  \fontdimen4\font\relax}
\providecommand{\BIBforeignlanguage}[2]{{%
\expandafter\ifx\csname l@#1\endcsname\relax
\typeout{** WARNING: IEEEtran.bst: No hyphenation pattern has been}%
\typeout{** loaded for the language `#1'. Using the pattern for}%
\typeout{** the default language instead.}%
\else
\language=\csname l@#1\endcsname
\fi
#2}}
\providecommand{\BIBdecl}{\relax}
\BIBdecl

\bibitem{chen:2018:yoga}
H.-T. Chen, Y.-Z. He, and C.-C. Hsu, ``Computer-assisted yoga training
  system,'' \emph{Multimedia Tools and Applications}, vol.~77, no.~18, pp.
  23\,969--23\,991, 2018.

\bibitem{clarke:2020:ReactV}
C.~Clarke, D.~Cavdir, P.~Chiu, L.~Denoue, and D.~Kimber, ``Reactive video:
  Adaptive video playback based on user motion for supporting physical
  activity,'' in \emph{Proceedings of the 33rd Annual ACM Symposium on User
  Interface Software and Technology}, 2020, pp. 196--208.

\bibitem{wang:2019:AICoach}
J.~Wang, K.~Qiu, H.~Peng, J.~Fu, and J.~Zhu, ``Ai coach: Deep human pose
  estimation and analysis for personalized athletic training assistance,'' in
  \emph{Proceedings of the 27th ACM International Conference on Multimedia},
  2019, pp. 374--382.

\bibitem{romanov:2002:PM}
N.~S. Romanov, \emph{Dr. Nicholas Romanov's Pose Method of Running: A New
  Paradigm of Running}.\hskip 1em plus 0.5em minus 0.4em\relax Pose Tech Corp.,
  2002.

\bibitem{dreyer:2009:CR}
D.~Dreyer and K.~Dreyer, \emph{ChiRunning: A revolutionary approach to
  effortless, injury-free running}.\hskip 1em plus 0.5em minus 0.4em\relax
  Simon and Schuster, 2009.

\bibitem{gleicher:2011:CIV}
M.~Gleicher, D.~Albers, R.~Walker, I.~Jusufi, C.~D. Hansen, and J.~C. Roberts,
  ``Visual comparison for information visualization,'' \emph{Information
  Visualization}, vol.~10, no.~4, pp. 289--309, 2011.

\bibitem{CoachesEye:2011:CE}
TechSmith, ``Coach's eye,'' \url{https://www.coachseye.com}, 2011.

\bibitem{Hudl:2011:PAT}
H.~Techiques, ``Hudl: Performance analysis tools for sports teams and athletes
  at every level,'' \url{https://www.hudl.com/en_gb/}, 2007.

\bibitem{OnForm:2021:VAS}
OnForm, ``Video analysis for skill development in any sport,''
  \url{https://www.getonform.com/sports}, 2021.

\bibitem{tang:2015:physio}
R.~Tang, X.-D. Yang, S.~Bateman, J.~Jorge, and A.~Tang, ``Physio@home:
  Exploring visual guidance and feedback techniques for physiotherapy
  exercises,'' in \emph{Proceedings of the 33rd Annual ACM Conference on Human
  Factors in Computing Systems}, 2015, pp. 4123--4132.

\bibitem{Matthew:2015:SMPL}
M.~Loper, N.~Mahmood, J.~Romero, G.~Pons-Moll, and M.~J. Black, ``{SMPL}: A
  skinned multi-person linear model,'' \emph{ACM Trans. Graphics (Proc.
  SIGGRAPH Asia)}, vol.~34, no.~6, pp. 248:1--248:16, Oct. 2015.

\bibitem{MotionPro:2018:MP}
MotionPro, ``Motion analysis software for all sports,''
  \url{https://www.motionprosoftware.com/index.htm}, 2018.

\bibitem{Woundefinedniak:2021:MSC}
P.~W. Wo\'{z}niak, M.~Zbytniewska, F.~Kiss, and J.~Niess, ``Making sense of
  complex running metrics using a modified running shoe,'' in \emph{Proceedings
  of the 2021 CHI Conference on Human Factors in Computing Systems}.\hskip 1em
  plus 0.5em minus 0.4em\relax New York, NY, USA: Association for Computing
  Machinery, 2021.

\bibitem{Qualysis:2015:QLS}
Q.~System, ``A leading provider of precision motion capture and 3d positioning
  tracking system,''
  \url{https://www.qualisys.com/applications/sports/running/}, 2015.

\bibitem{Kinovea:2004:AMYV}
Kinovea, ``A microscope for your videos,'' \url{https://www.kinovea.org/},
  2012.

\bibitem{velloso:2013:ExpNovice}
E.~Velloso, A.~Bulling, and H.~Gellersen, ``Motionma: motion modelling and
  analysis by demonstration,'' in \emph{Proceedings of the SIGCHI Conference on
  Human Factors in Computing Systems}, 2013, pp. 1309--1318.

\bibitem{fieraru:2021:AIFit}
M.~Fieraru, M.~Zanfir, S.~C. Pirlea, V.~Olaru, and C.~Sminchisescu, ``Aifit:
  Automatic 3d human-interpretable feedback models for fitness training,'' in
  \emph{Proceedings of the IEEE/CVF Conference on Computer Vision and Pattern
  Recognition}, 2021, pp. 9919--9928.

\bibitem{Once:2019:OP}
Once, ``Basketball video analysis all-in-one program,''
  \url{https://once.de/basketball-video-analysis/}, 2019.

\bibitem{stein:2017:BIP}
M.~Stein, H.~Janetzko, A.~Lamprecht, T.~Breitkreutz, P.~Zimmermann,
  B.~Goldl{\"u}cke, T.~Schreck, G.~Andrienko, M.~Grossniklaus, and D.~A. Keim,
  ``Bring it to the pitch: Combining video and movement data to enhance team
  sport analysis,'' \emph{IEEE transactions on visualization and computer
  graphics}, vol.~24, no.~1, pp. 13--22, 2017.

\bibitem{stein:2018:TSA}
M.~Stein, T.~Breitkreutz, J.~Haussler, D.~Seebacher, C.~Niederberger,
  T.~Schreck, M.~Grossniklaus, D.~Keim, and H.~Janetzko, ``Revealing the
  invisible: Visual analytics and explanatory storytelling for advanced team
  sport analysis,'' in \emph{2018 International Symposium on Big Data Visual
  and Immersive Analytics (BDVA)}, pp. 1--9.

\bibitem{Chen:2021:ASV}
Z.~Chen, S.~Ye, X.~Chu, H.~Xia, H.~Zhang, H.~Qu, and Y.~Wu, ``Augmenting sports
  videos with viscommentator,'' \emph{IEEE Transactions on Visualization and
  Computer Graphics}, vol.~28, no.~1, pp. 824--834, 2022.

\bibitem{wang:2022:TTVA}
J.~Wang, J.~Ma, K.~Hu, Z.~Zhou, H.~Zhang, X.~Xie, and Y.~Wu, ``Tac-trainer: A
  visual analytics system for iot-based racket sports training,'' \emph{IEEE
  Transactions on Visualization and Computer Graphics}, 2022.

\bibitem{lin:2022:QOE}
T.~Lin, Z.~Chen, Y.~Yang, D.~Chiappalupi, J.~Beyer, and H.~Pfister, ``The quest
  for omnioculars: Embedded visualization for augmenting basketball game
  viewing experiences,'' \emph{To appear in IEEE Transactions on Visualization
  and Computer Graphics}, 2022.

\bibitem{Semeraro:2022:VIP}
A.~Semeraro and L.~Turmo~Vidal, ``Visualizing instructions for physical
  training: Exploring visual cues to support movement learning from
  instructional videos,'' in \emph{Proceedings of the 2022 CHI Conference on
  Human Factors in Computing Systems}.\hskip 1em plus 0.5em minus 0.4em\relax
  New York, NY, USA: Association for Computing Machinery, 2022.

\bibitem{ortiz:2022:SVB}
V.~E. Ortiz-Padilla, M.~A. Ram{\'\i}rez-Moreno, G.~Presb{\'\i}tero-Espinosa,
  R.~A. Ram{\'\i}rez-Mendoza, and J.~d.~J. Lozoya-Santos, ``Survey on
  video-based biomechanics and biometry tools for fracture and injury
  assessment in sports,'' \emph{Applied Sciences}, vol.~12, no.~8, p. 3981,
  2022.

\bibitem{barre:2014:BT}
A.~Barre and S.~Armand, ``Biomechanical toolkit: Open-source framework to
  visualize and process biomechanical data,'' \emph{Computer methods and
  programs in biomedicine}, vol. 114, no.~1, pp. 80--87, 2014.

\bibitem{SkillSpector:2010:SS}
Video4coach, ``Video based motion and skill analysis,''
  \url{https://skillspector.apponic.com/}, 2010.

\bibitem{bartlett:2014:IntroSpBio}
R.~Bartlett, \emph{Introduction to sports biomechanics: Analysing human
  movement patterns}.\hskip 1em plus 0.5em minus 0.4em\relax Routledge, 2014.

\bibitem{wu:2022:IPM}
J.~Wu, D.~Liu, Z.~Guo, and Y.~Wu, ``Rasipam: Interactive pattern mining of
  multivariate event sequences in racket sports,'' \emph{IEEE Transactions on
  Visualization and Computer Graphics}, 2022.

\bibitem{su:2018:ISN}
W.~Su, D.~Du, X.~Yang, S.~Zhou, and H.~Fu, ``Interactive sketch-based normal
  map generation with deep neural networks,'' \emph{Proceedings of the ACM on
  Computer Graphics and Interactive Techniques}, vol.~1, no.~1, pp. 1--17,
  2018.

\bibitem{su:2022:DIS}
W.~Su, H.~Ye, S.-Y. Chen, L.~Gao, and H.~Fu, ``Drawinginstyles: Portrait image
  generation and editing with spatially conditioned stylegan,'' \emph{IEEE
  Transactions on Visualization and Computer Graphics}, 2022.

\bibitem{Nebeling:2015:KA}
M.~Nebeling, D.~Ott, and M.~C. Norrie, ``Kinect analysis: a system for
  recording, analysing and sharing multimodal interaction elicitation
  studies,'' in \emph{Proceedings of the 7th ACM SIGCHI Symposium on
  Engineering Interactive Computing Systems}, 2015, pp. 142--151.

\bibitem{MSkinect:2017:VGB}
Microsoft, ``Visual gesture builder (vgb),''
  \url{https://thewindowsupdate.com/2017/05/04/visual-gesture-builder-vgb/},
  2017.

\bibitem{mathis:2018:deeplabcut}
A.~Mathis, P.~Mamidanna, K.~M. Cury, T.~Abe, V.~N. Murthy, M.~W. Mathis, and
  M.~Bethge, ``Deeplabcut: markerless pose estimation of user-defined body
  parts with deep learning,'' \emph{Nature neuroscience}, vol.~21, no.~9, pp.
  1281--1289, 2018.

\bibitem{anderson:2013:youmove}
F.~Anderson, T.~Grossman, J.~Matejka, and G.~Fitzmaurice, ``Youmove: enhancing
  movement training with an augmented reality mirror,'' in \emph{Proceedings of
  the 26th annual ACM symposium on User interface software and technology},
  2013, pp. 311--320.

\bibitem{Suzuki:2020:ERG}
R.~Suzuki, R.~H. Kazi, L.-y. Wei, S.~DiVerdi, W.~Li, and D.~Leithinger,
  ``Realitysketch: Embedding responsive graphics and visualizations in ar
  through dynamic sketching,'' in \emph{Proceedings of the 33rd Annual ACM
  Symposium on User Interface Software and Technology}, 2020, pp. 166--181.

\bibitem{hay:1978:BST}
J.~Hay, \emph{The biomechanics of sports techniques}.\hskip 1em plus 0.5em
  minus 0.4em\relax Prentice-Hall, 1978.

\bibitem{wulf:2010:EVI}
G.~Wulf, C.~Shea, and R.~Lewthwaite, ``Motor skill learning and performance: a
  review of influential factors,'' \emph{Medical education}, vol.~44, no.~1,
  pp. 75--84, 2010.

\bibitem{shneiderman:2003:EHI}
B.~Shneiderman, ``The eyes have it: A task by data type taxonomy for
  information visualizations,'' in \emph{The craft of information
  visualization}.\hskip 1em plus 0.5em minus 0.4em\relax Elsevier, 2003, pp.
  364--371.

\bibitem{novacheck:1998:BR}
T.~F. Novacheck, ``The biomechanics of running,'' \emph{Gait \& posture},
  vol.~7, no.~1, pp. 77--95, 1998.

\bibitem{van:2021:BRS}
B.~T. van Oeveren, C.~J. de~Ruiter, P.~J. Beek, and J.~H. van Die{\"e}n, ``The
  biomechanics of running and running styles: a synthesis,'' \emph{Sports
  biomechanics}, pp. 1--39, 2021.

\bibitem{choi:2021:TCMR}
H.~Choi, G.~Moon, J.~Y. Chang, and K.~M. Lee, ``Beyond static features for
  temporally consistent 3d human pose and shape from a video,'' in
  \emph{Proceedings of the IEEE/CVF Conference on Computer Vision and Pattern
  Recognition}, 2021, pp. 1964--1973.

\bibitem{holden:2017:LMP}
D.~Holden, T.~Komura, and J.~Saito, ``Phase-functioned neural networks for
  character control,'' \emph{ACM Transactions on Graphics (TOG)}, vol.~36,
  no.~4, pp. 1--13, 2017.

\bibitem{berndt:1994:DTW}
D.~J. Berndt and J.~Clifford, ``Using dynamic time warping to find patterns in
  time series.'' in \emph{KDD workshop}, vol.~10, no.~16.\hskip 1em plus 0.5em
  minus 0.4em\relax Seattle, WA, USA:, 1994, pp. 359--370.

\bibitem{shu:2020:GIF}
X.~Shu, A.~Wu, J.~Tang, B.~Bach, Y.~Wu, and H.~Qu, ``What makes a data-gif
  understandable?'' \emph{IEEE Transactions on Visualization and Computer
  Graphics}, vol.~27, no.~2, pp. 1492--1502, 2020.

\bibitem{lee:2005:MSY}
C.~H. Lee, A.~Varshney, and D.~W. Jacobs, ``Mesh saliency,'' in \emph{ACM
  SIGGRAPH 2005 Papers}, 2005, pp. 659--666.

\bibitem{fu:2008:UOM}
H.~Fu, D.~Cohen-Or, G.~Dror, and A.~Sheffer, ``Upright orientation of man-made
  objects,'' in \emph{ACM SIGGRAPH 2008 papers}, 2008, pp. 1--7.

\bibitem{Chen:2014:HVA}
H.-T. Chen, T.~Grossman, L.-Y. Wei, R.~M. Schmidt, B.~Hartmann, G.~Fitzmaurice,
  and M.~Agrawala, ``History assisted view authoring for 3d models,'' in
  \emph{Proceedings of the SIGCHI Conference on Human Factors in Computing
  Systems}, 2014, pp. 2027--2036.

\bibitem{Kiciroglu:2020:AMC}
S.~Kiciroglu, H.~Rhodin, S.~N. Sinha, M.~Salzmann, and P.~Fua, ``Activemocap:
  Optimized viewpoint selection for active human motion capture,'' in
  \emph{Proceedings of the IEEE/CVF Conference on Computer Vision and Pattern
  Recognition}, 2020, pp. 103--112.

\bibitem{Kwon:2022:OCP}
B.~Kwon, J.~Huh, K.~Lee, and S.~Lee, ``Optimal camera point selection toward
  the most preferable view of 3-d human pose,'' \emph{IEEE Transactions on
  Systems, Man, and Cybernetics: Systems}, vol.~52, no.~1, pp. 533--553, 2022.

\bibitem{mirzabekiantz:2016:BMN}
E.~Mirzabekiantz, ``Benesh movement notation for humanoid robots?'' in
  \emph{Dance Notations and Robot Motion}.\hskip 1em plus 0.5em minus
  0.4em\relax Springer, 2016, pp. 299--317.

\bibitem{brooke:1996:SUS}
J.~Brooke \emph{et~al.}, ``Sus-a quick and dirty usability scale,''
  \emph{Usability evaluation in industry}, vol. 189, no. 194, pp. 4--7, 1996.

\bibitem{romanov:2002:PMR}
N.~S. Romanov, \emph{Dr. Nicholas Romanov's Pose method of running: A new
  paradigm of running}.\hskip 1em plus 0.5em minus 0.4em\relax Pose Tech Corp.,
  2002.

\bibitem{chu:2021:TIVEE}
X.~Chu, X.~Xie, S.~Ye, H.~Lu, H.~Xiao, Z.~Yuan, Z.~Chen, H.~Zhang, and Y.~Wu,
  ``Tivee: Visual exploration and explanation of badminton tactics in immersive
  visualizations,'' \emph{IEEE Transactions on Visualization and Computer
  Graphics}, vol.~28, no.~1, pp. 118--128, 2021.

\bibitem{williams:1987:RE}
K.~R. Williams and P.~R. Cavanagh, ``Relationship between distance running
  mechanics, running economy, and performance,'' \emph{Journal of Applied
  Physiology}, vol.~63, no.~3, pp. 1236--1245, 1987.

\bibitem{aristidou:2018:DMS}
A.~Aristidou, D.~Cohen-Or, J.~K. Hodgins, Y.~Chrysanthou, and A.~Shamir, ``Deep
  motifs and motion signatures,'' \emph{ACM Transactions on Graphics (TOG)},
  vol.~37, no.~6, pp. 1--13, 2018.

\bibitem{hanley:2018:DMV}
B.~Hanley, C.~B. Tucker, and A.~Bissas, ``Differences between motion capture
  and video analysis systems in calculating knee angles in elite-standard race
  walking,'' \emph{Journal of sports sciences}, vol.~36, no.~11, pp.
  1250--1255, 2018.

\bibitem{reinking:2018:RTD}
M.~F. Reinking, L.~Dugan, N.~Ripple, K.~Schleper, H.~Scholz, J.~Spadino,
  C.~Stahl, and T.~G. McPoil, ``Reliability of two-dimensional video-based
  running gait analysis,'' \emph{International Journal of Sports Physical
  Therapy}, vol.~13, no.~3, p. 453, 2018.

\bibitem{pipkin:2016:RVA}
A.~Pipkin, K.~Kotecki, S.~Hetzel, and B.~Heiderscheit, ``Reliability of a
  qualitative video analysis for running,'' \emph{journal of orthopaedic \&
  sports physical therapy}, vol.~46, no.~7, pp. 556--561, 2016.

\bibitem{kuhtz:2012:AUL}
J.~P. Kuhtz-Buschbeck and B.~Jing, ``Activity of upper limb muscles during
  human walking,'' \emph{Journal of Electromyography and Kinesiology}, vol.~22,
  no.~2, pp. 199--206, 2012.

\bibitem{hinrichs:1990:WBM}
R.~N. Hinrichs, ``Whole body movement: coordination of arms and legs in walking
  and running,'' \emph{Multiple muscle systems}, pp. 694--705, 1990.

\bibitem{wei:2021:DPA}
R.~X. Wei, Z.~Y. Chan, J.~H. Zhang, G.~L. Shum, C.-Y. Chen, and R.~T. Cheung,
  ``Difference in the running biomechanics between preschoolers and adults,''
  \emph{Brazilian Journal of Physical Therapy}, vol.~25, no.~2, pp. 162--167,
  2021.

\bibitem{zecha:2018:SJF}
D.~Zecha, C.~Eggert, M.~Einfalt, S.~Brehm, and R.~Lienhart, ``A convolutional
  sequence to sequence model for multimodal dynamics prediction in ski jumps,''
  in \emph{Proceedings of the 1st International Workshop on Multimedia Content
  Analysis in Sports}, 2018, pp. 11--19.

\bibitem{Seth:2018:OS}
A.~Seth, J.~L. Hicks, T.~K. Uchida, A.~Habib, C.~L. Dembia, J.~J. Dunne, C.~F.
  Ong, M.~S. DeMers, A.~Rajagopal, M.~Millard \emph{et~al.}, ``Opensim:
  Simulating musculoskeletal dynamics and neuromuscular control to study human
  and animal movement,'' \emph{PLoS computational biology}, vol.~14, no.~7, p.
  e1006223, 2018.

\end{thebibliography}


\vspace{-10mm}
\begin{IEEEbiography}[{\includegraphics[width=1in,height=1.25in,clip,keepaspectratio]{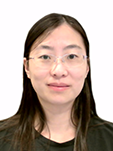}}]
{Jingyuan Liu}{\,} is PhD candidate in the Department of Computer Science and Engineering at the Hong Kong University of Science and Technology. Before joining HKUST, she received a Master degree in Control Science and Engineering from Beihang University and a BENG degree in Automation from Beijing University of Science and Technology. Her research interests are in video analysis and user interface design.
\end{IEEEbiography}

\vspace{-12mm}
\begin{IEEEbiography}[{\includegraphics[width=1in,height=1.25in,clip,keepaspectratio]{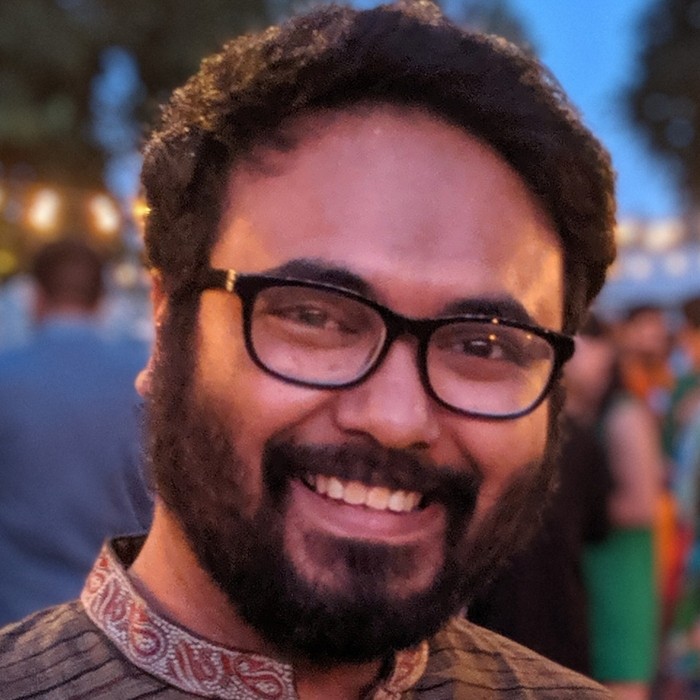}}]
{Nazmus Saquib}{\,} is currently the CTO of Kickstarter's Creative Crowdfunding Research and Protocol labs, focused on building tools for technologists and creative artists. His academic research spans several areas in visual computing and social computing. Previously, he founded Universal Machine Inc., a technical venture studio in Silicon Valley that has incubated startups in transportation, agriculture, climate, and media. Saquib obtained his PhD from MIT Media Lab, researching and building tools for embodied cognition, mathematics, and sensor technology. He has an MS in Computational Engineering and a BA in Physics.
\end{IEEEbiography}

\begin{IEEEbiography}[{\includegraphics[width=1in,height=1.25in,clip,keepaspectratio]{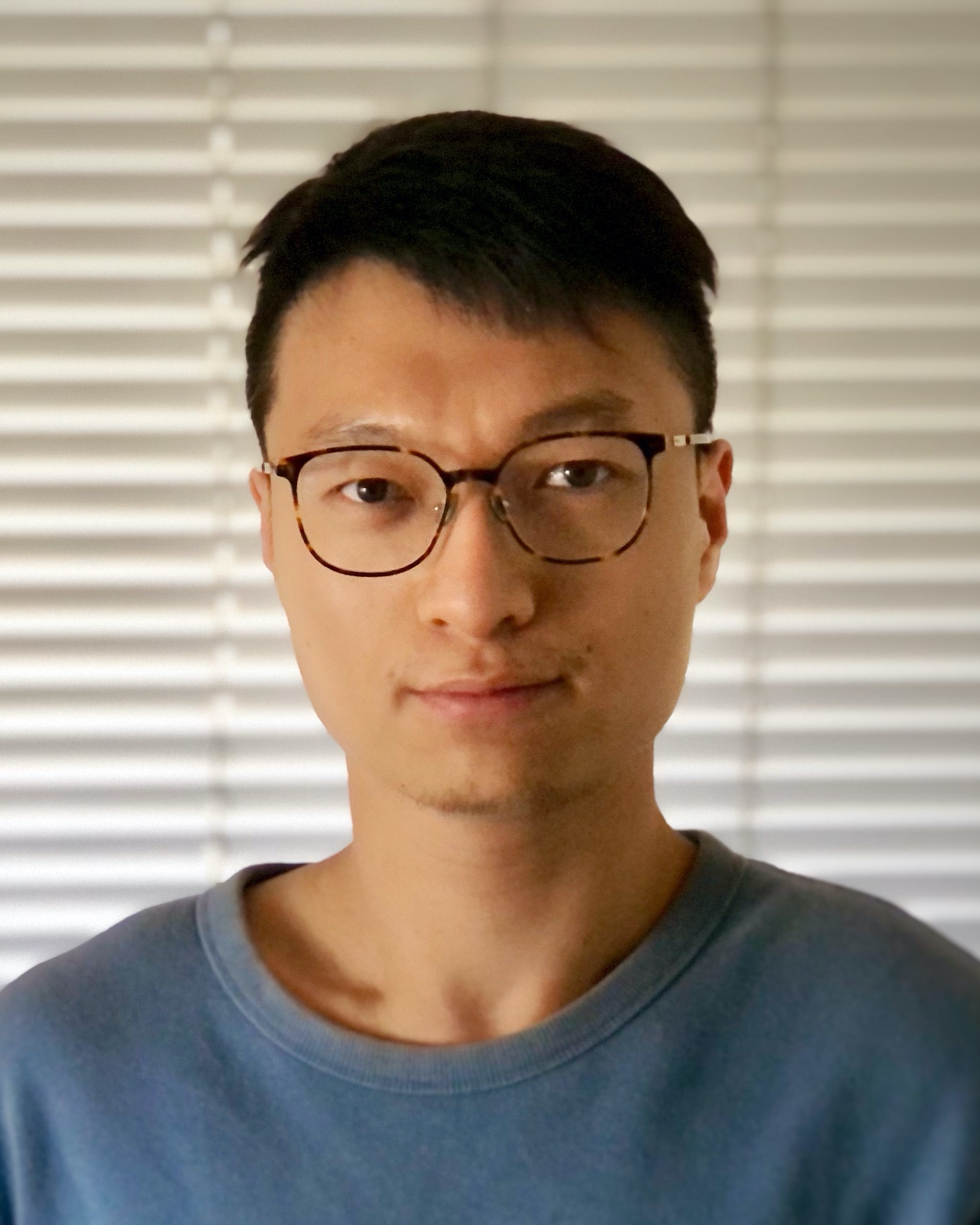}}]
{Zhutian Chen}{\,} is a postdoc at the Visual Computing Group at Harvard University. Before joining Harvard, he was a postdoc at University of California San Diego, and a Ph.D. student at Hong Kong University of Science and Technology. His interests are in Information Visualization, Human-Computer Interaction, and Augmented Reality.
\end{IEEEbiography}
\vspace{-9mm}

\begin{IEEEbiography}[{\includegraphics[width=1in,height=1.25in,clip,keepaspectratio]{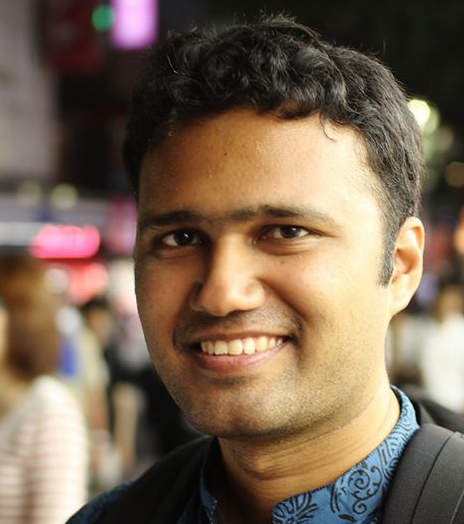}}]
{Rubaiat Habib Kazi}{\,} is a Sr. Research Scientist at Adobe Research. He designs and develops computing tools that facilitate powerful ways of thinking, design, and communication with sketching and gestures. His research in animation and dynamic drawings is turned into new products that reach to a global audience. He also contributed to the design and engineering of the inaugural animation/motion features in Adobe Fresco. Prior to Adobe, He worked at Autodesk Research, Microsoft Research, and Japan Science and Technology Agency.
\end{IEEEbiography}
\vspace{-9mm}

\begin{IEEEbiography}[{\includegraphics[width=1in,height=1.25in,clip,keepaspectratio]{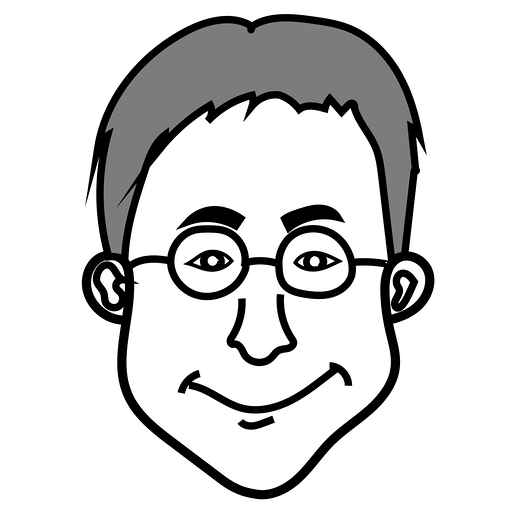}}]
{Li-Yi Wei}{\,} is a research scientist with Adobe Research, and has been a professor with the University of Hong Kong, a researcher with Microsoft Research, and a GPU architect with NVIDIA. He received his PhD degree from Stanford University. He has served as associate editors for ACM ToG and EG CGF, the top two journals in computer graphics and interactive techniques, advisory board member for PACMCGIT, a guest editor for IEEE TVCG, paper co-chairs for EGSR and I3D, and paper committee members for SIGGRAPH, SIGGRAPH Asia, and EG. He has published 40+ papers in top graphics/HCI venues such as ToG/SIGGRAPH/CHI/UIST and received 20+ patents. As a primary adviser or research mentor, he has guided 10+ students to become researchers/engineers in top companies or professors in top schools.
\end{IEEEbiography}
\vspace{-9mm}

\begin{IEEEbiography}[{\includegraphics[width=1in,height=1.25in,clip,keepaspectratio]{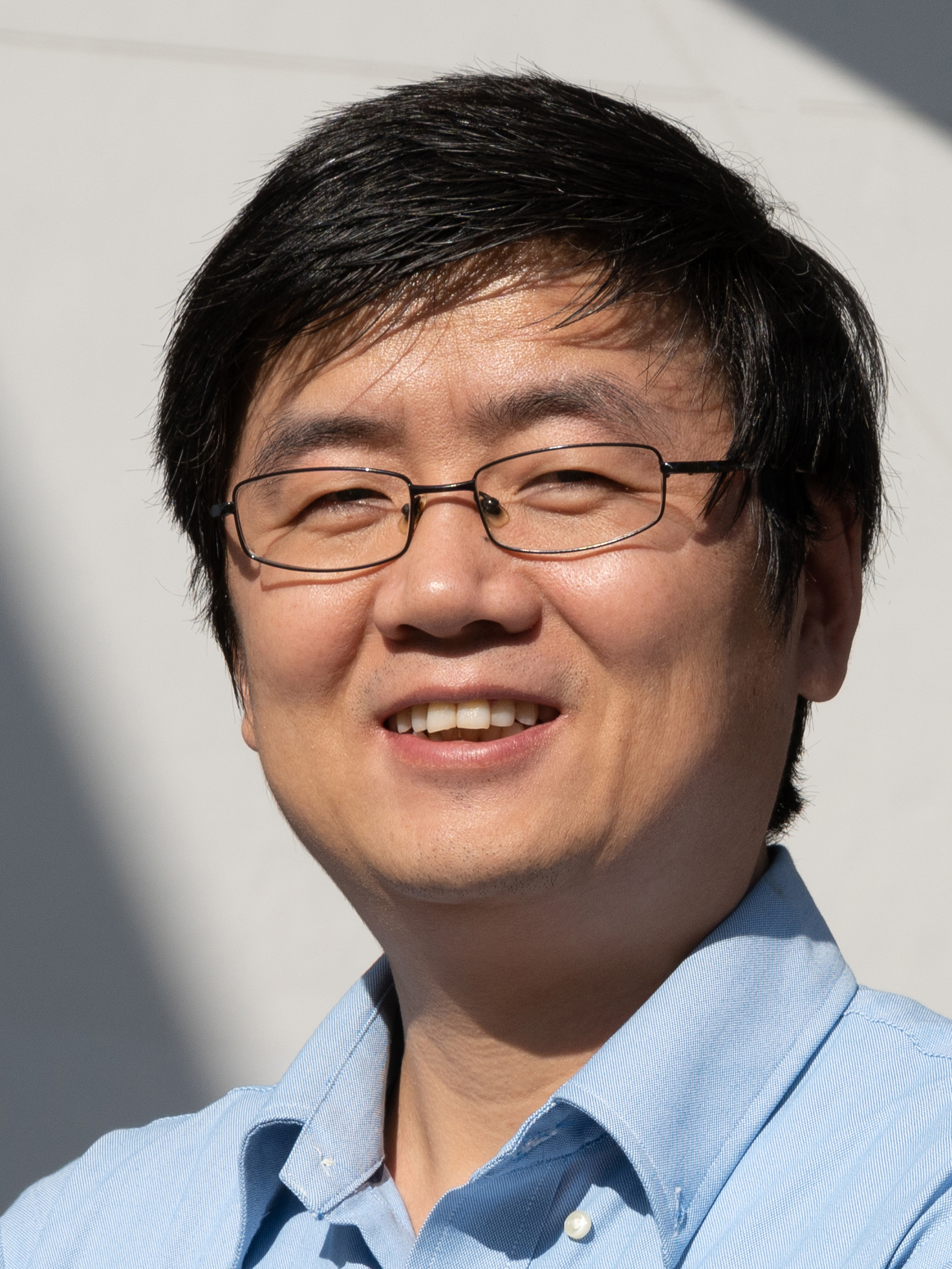}}]
{Hongbo Fu}{\,} is a Professor in the School of Creative Media, City University of Hong Kong. Before joining CityU, he had postdoctoral research training at the Imager Lab, University of British Columbia, Canada, and the Department of Computer Graphics, Max-Planck-Institut Informatik, Germany. He received a PhD degree in computer science from the Hong Kong University of Science and Technology in 2007 and a BS degree in information sciences from Peking University, China, in 2002. His primary research interests fall in the fields of computer graphics and human-computer interaction. He has served as an associate editor of The Visual Computer, Computers\&Graphics, and Computer Graphics Forum.
\end{IEEEbiography}
\vspace{-9mm}

\begin{IEEEbiography}[{\includegraphics[width=1in,height=1.25in,clip,keepaspectratio]{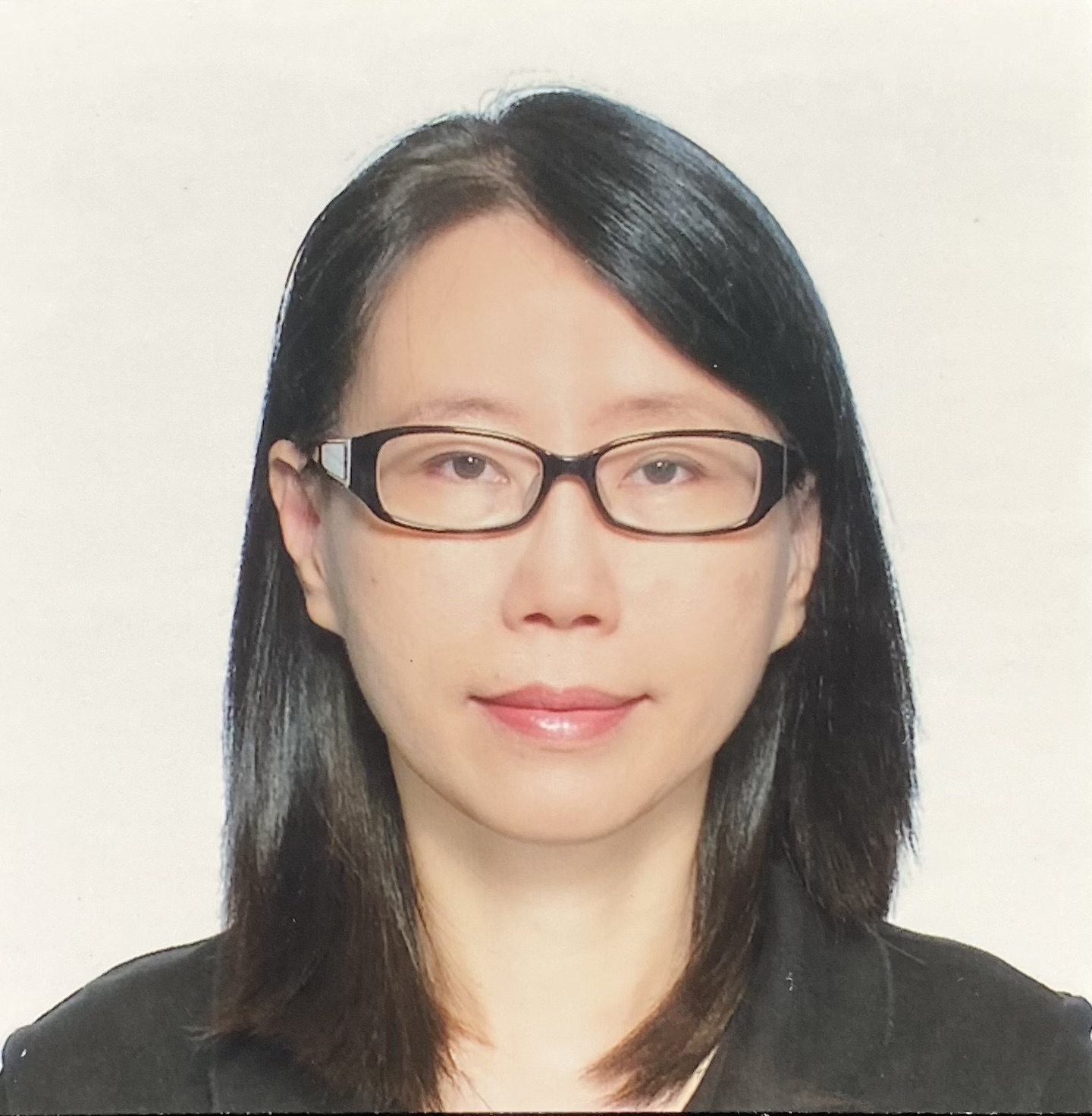}}]
{Chiew-Lan Tai}{\,} is a Professor at the Department of Computer Science and Engineering in the Hong Kong University of Science and Technology.
She received a D.Sc degree in Information Science from the University of Tokyo in 1997, a M.Sc. degree in Computer and Information Sciences from the National University of Singapore in 1990, and a B.Sc. degree in Mathematics from the University of Malaya in 1985.  Her research interests include digital geometry processing, computer graphics, and computer vision.
\end{IEEEbiography}

\begin{comment}
\begin{IEEEbiographynophoto}{John Doe}
Biography text here.
\end{IEEEbiographynophoto}

%\newpage

\begin{IEEEbiographynophoto}{Jane Doe}
Biography text here.
\end{IEEEbiographynophoto}
\end{comment}


%\vfill


\end{document}